\documentclass[a4paper,10pt]{article}
\usepackage{jheppub} 
\usepackage{natbib}
\usepackage{orcidlink}
\newcommand{\nn}{\nonumber}
\usepackage{amsmath,amssymb}
\usepackage[toc,page]{appendix}
\usepackage{comment}
\usepackage{hyperref}
\usepackage{mathrsfs}
\usepackage{mathtools}
\usepackage{empheq}
\usepackage{academicons}
\title{\boldmath Light and Shadow OPEs: A Carroll Symmetric Approach to Flat Holography}

\author[1]{Sourish Banerjee \orcidlink{0000-0003-2618-1025}}
\author[1]{Rudranil Basu \orcidlink{0000-0003-0655-0890}}

\affiliation[1]{Department of Physics, Birla Institute of Technology and Science, Pilani, K K Birla Goa Campus, Zuarinagar, Sancoale, Goa 403726, India}

\emailAdd{p20210001@goa.bits-pilani.ac.in}
\emailAdd{rudranilb@goa.bits-pilani.ac.in}

\abstract{In this work, we fix the leading term in the operator algebra of light-transformed operators in the context of flat space holography. Starting with light-transformed graviton correlators, we show that the OPE obtained by taking the collinear limit satisfies translation symmetry at the leading order not independently but with assistance from the sub-leading order. Motivated by this result, we start with a general CFT-like ansatz for the OPE and, by keeping track of the sub-leading term, derive an expression for the scaling dimension of the operator appearing in the leading term. We further show that we can fix the OPE coefficient corresponding to this term as well. We use this result to find the OPE for well-studied theories like gravity, Yang-Mills and Einstein-Yang-Mills theory, and find that the scaling dimension found from our formula matches with the one obtained by looking at the collinear limit of bulk momentum space vertices. We initiate a similar study for operators in the shadow basis by looking at the OPE of shadow-transformed graviton correlators.}

\begin{document}
\maketitle
\flushbottom
 \section{Introduction}
The correspondence between a bulk theory of gravity and a conformal field theory defined on the boundary of an anti-de Sitter spacetime is very well studied \cite{Witten:1998qj, Maldacena:1997re}. The advantage of having a CFT dual to a bulk theory is that one can use conformal symmetry to fix the spectrum of the CFT, which then can be used to calculate local observables, by which we mean correlation functions of local operators. To be more precise, on the CFT side, one writes down the algebra of the operators and tries to constrain its form using conformal symmetry. The algebra of operators or the operator product expansion, for two operators with conformal weights $(h_{1},\bar{h}_{1})$ and $(h_{2},\bar{h}_{2})$ takes the form \cite{DiFrancesco:1997nk}, 
\begin{equation}
    \mathcal{O}_{1}\mathcal{O}_{2}=\sum_{p,K.\bar{K}}C^{K,\bar{K}}_{12p}z_{12}^{h_{p}-h_{1}-h_{2}+K}\bar{z}_{12}^{\bar{h}_{p}-\bar{h}_{1}-\bar{h}_{2}+\bar{K}}\partial_{z_{2}}^{K}\partial_{\bar{z}_{2}}^{\bar{K}}\mathcal{O}_{p} \label{formalOPE}
\end{equation}
where the power law behaviour of $z_{12}$ and $\bar{z}_{12}$ is fixed using conformal symmetry. Now, using such an expansion, one can calculate all higher-point correlation functions starting with the two-point function provided one knows the OPE coefficient $C^{K,\bar{K}}_{12p}$ and the weight of the conformal primary $\mathcal{O}_{p}$, $(h_{p},\bar{h}_{p})$. Quite rightly, these two quantities are known as CFT data. This idea of calculating correlation functions using the OPE of local operators in a CFT is known as the bootstrap program \cite{Poland:2018epd, Simmons-Duffin:2016gjk, Rychkov:2016iqz}. Thus, using the bulk-boundary correspondence, one can now map the observables calculated in the CFT to observables in the bulk, leading to information about the physics of the bulk.

The past decade has seen considerable progress in the field of flat-space holography with two equivalent approaches, namely Celestial CFT and Carrollian CFT. In celestial CFT, motivated by the fact that the Lorentz transformations act as global conformal transformations of the Riemann sphere, one maps $4D$ momentum space quantum field theory to a CFT defined on a codimension two celestial sphere at null infinity \cite{Pasterski:2016qvg, Pasterski:2017ylz, Strominger:2017zoo, Donnay:2023mrd}. This map is achieved by performing the Mellin transform of momentum modes, based on which it was shown that the tree-level scattering amplitude of massless particles is mapped to correlation functions of primary operators inserted on the celestial sphere. Owing to momentum conservation in the bulk, the lower-point correlation functions turn out to be ultra-local. One of the obvious next steps in this program was to study the algebra of these operators, which was done by Mellin transforming the collinear singularities of bulk scattering amplitudes \cite{Pate:2019lpp, Guevara:2021abz}. In this line of work, the scaling dimension of the leading term in the OPE was fixed by looking at the mass dimension of the bulk three-point vertex from which the collinear limit was taken, and the OPE coefficient was fixed by imposing the constraint that the OPE respect the translation and asymptotic symmetries. This approach has taught us a lot about the algebra of operators of the putative dual CFT living on the celestial sphere \cite{Himwich:2021dau, Strominger:2021mtt, Surubaru:2025qhs, Bhattacharyya:2025nfp, Ren:2023trv, McLoughlin:2024ldp}. However, the approach of extracting information about bulk physics by studying only the boundary CFT in its own right warrants due attention. 

This brings us to the second avenue towards finding a theory dual to $4D$ flat spacetime, which is motivated by the fact that the symmetry algebra of $4D$ asymptotically flat spacetime, the BMS symmetries, is isomorphic to the conformal Carroll algebra in $3D$ \cite{Barnich:2006av,bagchi2010correspondence, Bagchi:2012cy, Duval:2014uva}\footnote{We point the interested reader to \cite{Bagchi:2025vri} for a concise review on the latest developments in Carrollian field theories.}. Parallely, it was found that the correlation function of gravitons in the Mellin basis was divergent, a possible solution to this problem, amongst others \cite{Stieberger:2018edy, Mitra:2024ugt}, was to introduce a regulator in the definition of the Mellin transform \cite{Banerjee:2019prz}. The introduction of this regulator appended the Mellin transformed operator with BMS symmetries \cite{Banerjee:2020kaa}. This persuaded one to look at theories with conformal Carroll symmetry. Such theories were pioneered in \cite{Bagchi:2022emh} by performing a modified Mellin transform of the bulk momentum modes \cite{Banerjee:2018gce, Banerjee:2020kaa, Banerjee:2019prz}, which is nothing but a time translation of the usual Mellin operator along the null-time direction on $\mathcal{I}^{\pm}$ and in \cite{Donnay:2022aba, Donnay:2022wvx} wherein the authors used the Kirchhoff d'Adh\'emar formula to obtain operators of the boundary theory from bulk. The possible functional forms of correlation functions of conformal Carroll operators were found in \cite{Nguyen:2023miw} and these were independently obtained by performing the aforementioned integral transforms in \cite{Mason:2023mti} and by considering appropriate limits of AdS Witten diagrams in \cite{Bagchi:2023cen, Bagchi:2023fbj, Alday:2024yyj, Surubaru:2025fmg, Kraus:2024gso, Chakrabortty:2024bvm}. Significant progress has been made in understanding conformal Carrollian theories from an intrinsic perspective as well, stress tensor OPEs for scalars was calculated in \cite{Dutta:2022vkg} and in \cite{Saha:2023hsl, Saha:2023abr, Ruzziconi:2024kzo, Bagchi:2024gnn, Agrawal:2024sju} the authors were able to derive the Ward identities corresponding to the leading, sub-leading soft graviton theorem and they could derive the $w_{1+\infty}$\cite{Strominger:2021mtt} algebra of soft gravitons from a purely Carrollian consideration. In \cite{Mason:2023mti, Banerjee:2020kaa}, the authors wrote down the OPE of conformal Carroll operators, obtained by performing a modified Mellin transform, by taking the collinear limit of the Carrollian correlation functions thus obtained. This again required input from the bulk side. Recently, in \cite{Nguyen:2025sqk}, the possible structure of Carrollian OPEs was derived, and the OPEs were classified by whether one arrives at them by taking the ``uniform coincidence limit" or the ``holomorphic coincidence limit". However, as is expected in Carrollian theories \cite{Ecker:2024czx,Sharma:2025rug, Ara:2024vbe, Chen:2024voz, Cotler:2025dau, Cotler:2024xhb, Banerjee:2023jpi, Liu:2024nfc}, the correlation functions are ultra-local. Consequently, the OPEs constructed out of them had contact terms which, from the perspective of ordinary CFT, are not considered a part of the OPE. This is a hindrance if one wants to construct a bootstrap program for Carroll CFT because then the OPEs would not be able to reproduce all correlation functions. 

The primary objective of trying to find a basis where correlation functions of either celestial or Carrollian CFT are not ultra-local is to make the theory more akin to usual CFT so that the robust techniques of CFT can be used further to study the boundary theory from an intrinsic perspective. This can be achieved in multiple ways: a) analytically continue to (2,2) signature and work with celestial leaf amplitudes \cite{Melton:2023bjw, Melton:2024jyq, Mandal:2024zao}, b) study the theory in a non-trivial background or couple the bulk theory to massive scalar \cite{Ball:2023ukj, Ruzziconi:2024zkr, Fan:2022vbz} and c) change the basis of operators by performing an integral transform like the shadow transform \cite{Osborn:2012vt, Crawley:2021ivb, Chang:2022jut, Banerjee:2022wht} or the light transform \cite{Kravchuk:2018htv, Sharma:2021gcz, Banerjee:2022hgc, De:2022gjn, Hu:2022syq}. In our previous work \cite{Banerjee:2024hvb}, we had shown that if we perform a light transform of the modified Mellin operators, then the correlation functions obtained thereof are non-distributional in nature, which also satisfy the Ward identities of global BMS transformations. In addition to this, we found that the global part of supertranslations on $\mathcal{I}^{+}$ acts as derivatives along the sphere coordinates $(z,\bar{z})$. This enables one to fix the conformal weight of the operator appearing in the leading term in the OPE of gluons, along with the OPE coefficient without resorting to any bulk input. The result thus obtained matches the result one gets by taking a collinear limit of the light-transformed three-point function. This was a crucial result since it implies that we can fix part of the CFT data (data corresponding to the leading term only) for the Carroll CFT dual to $4D$ flat spacetime. This takes us a step towards bootstrapping the dual boundary CFT. 

Further proceeding in the above-mentioned direction, we extend the results obtained for light-transformed gluon operators to light-transformed conformal Carroll primaries of any spin in this work. We also show that one can indeed fix the leading term in the OPE. However, unlike our previous work, we encountered a technical subtlety here. To fix the leading term, one has to consider the contribution from the sub-leading, which was not the case for gluons. We show that our formula for calculating the scaling dimension, equation \eqref{generalScaling}, gives the correct expression for the scaling dimension of gravitons, gluons, and gluons coupled to gravitons. Consequently, we can infer information about bulk physics, specifically the dimension of the vertex and the type of particles interacting via that vertex\cite{Pate:2019lpp}, by symmetry considerations in the boundary theory. Recently, in \cite{Narayanan:2024qgb}, the author found that one can construct marginal gluon operators using light transformation, which in turn gives us a better understanding of the dual CFT living on the boundary. We consider the OPE of such marginal operators as well and find that this OPE receives a contribution from a light-transformed conformally soft gluon operator. 

A natural question to ask now is whether one can extend such an analysis to the space of shadow-transformed operators. Lately, there has been some work on understanding the role of shadow transform in flat space holography in \cite{Banerjee:2022wht, Banerjee:2024yir} and especially in \cite{Himwich:2025bza} where the authors have initiated a study on the algebra of shadow operators in celestial CFT. In this work, we find the OPE of shadow-transformed graviton primaries and check that it satisfies translation symmetry as well. Formulating an intrinsic approach to study the OPE of shadow primaries is a goal for future work.

The rest of the paper is organised as follows: In section \ref{Graviton OPE in light-transformed basis}, we find the graviton OPE by taking the collinear limit of the light-transformed graviton anti-MHV correlator in Carrollian basis. We further check whether the OPE is consistent with translation symmetry, we act on it with $\delta_{M_{00}}$ symmetry and show that indeed the OPE is invariant under translations. Following that, in section \ref{General structure of the operator product expansion}, we write down a general formula for the scaling dimension of the operator contributing to the leading term in the OPE. Using this, in section \ref{Fixing the leading OPE coefficient}, we find a set of equations that one can solve to find the OPE coefficients. We employ these relations to calculate the leading term of the OPE of gravitons and find that they match with the results of section \ref{Graviton OPE in light-transformed basis}. We also calculate OPEs of gluons and gravitons coupled to gluons. In section \ref{Shadow graviton OPE} we commence our study of the operator algebra pertaining to the space of shadow-transformed operators by looking at the behaviour of shadow shadow-transformed graviton three-point function under the collinear limit. In section \ref{GlobalBMSShadow} we perform a symmetry analysis of the OPE thus obtained and verify that the OPE respects the symmetries of the theory.

\section{Graviton OPE in light-transformed basis} \label{Graviton OPE in light-transformed basis}
The light transform, originally defined in \cite{Kravchuk:2018htv}, was designed to construct non-local operators with continuous spin. Motivated by this definition, we define the light transform as a contour integral in the following manner \cite {Banerjee:2022hgc},
\begin{align}
    & L^{+}[\phi_{a,\mathfrak{h}}](u,z,\Bar{z})=\oint_{w=z} dw \ \frac{\phi_{a,\mathfrak{h}}(u,w,\Bar{z})}{(w-z)^{2-2h}} \label{L+} \\
    & L^{-}[\phi_{a,\mathfrak{h}}](u,z,\Bar{z})=\oint_{\Bar{w}=\Bar{z}} d\Bar{w} \ \frac{\phi_{a,\mathfrak{h}}(u,z,\Bar{w})}{(\Bar{w}-\Bar{z})^{2-2\Bar{h}}}\label{L-}
\end{align}
where the operator $\phi_{a,\mathfrak{h}}(u,z,\Bar{z})$ is obtained by performing a modified Mellin transformation of a momentum mode\cite{Banerjee:2018gce}. Here and in other parts of this work, we use $\mathfrak{h}=(h,\bar{h})$.
\subsection{Collinear limit} \label{SectionCollinearLimit}
In this section, we obtain the OPE of light-transformed graviton operators by taking the collinear limit of the three-point anti-MHV graviton correlator written in light-transformed basis. Before we start the calculation for OPE from collinear limit, let us, for completeness, note down the two-point function in terms of gravitons,
\begin{align}
    \langle G^{+}_{\Delta_{1}}(z_1, \bar{z}_1, u_1)G^{-}_{\Delta_{2}}(z_2, \bar{z}_2, u_2)\rangle=\frac{\Gamma(\Delta_{1}+\Delta_{2}-2)}{(iu_{21})^{\Delta_{1}+\Delta_{2}-2}}\delta(z_{12})\delta(\bar{z}_{12}) .\label{CarrollGraviton2pt}
\end{align}
The anti-MHV graviton three-point correlator\footnote{We obtain this in the usual way of replacing $z$ with $\bar{z}$ in the three-point MHV graviton correlator of \cite{Bagchi:2023cen}.},
\begin{align}
     \langle G^{+}_{\Delta_{1}}G^{+}_{\Delta_{2}}G^{-}_{\Delta_{3}}\rangle=(-1)^{\Delta_{1}+\Delta_{2}}\Gamma(\beta-2) \bar{z}_{12}^{\Delta_{3}+2}\bar{z}_{23}^{\Delta_{1}-2}\bar{z}_{31}^{\Delta_{2}-2}\frac{\delta(z_{12})\delta(z_{23})}{[i(\bar{z}_{1}u_{32}+\bar{z}_{2}u_{13}+\bar{z}_{3}u_{21})]^{\beta-2}}, \label{Anti-MHGravitonMellin}
\end{align}
where $\beta=\sum_{i=1}^{3}\Delta_{i}$. Again, as stated before, one must consider the correlator of Carrollian gravitons since this correlator diverges in the ordinary Mellin basis. From this, using \eqref{L+}, \eqref{L-} we obtain the following two and three-point functions,
\begin{align}
    \langle L^{+}[G^{+}_{\Delta_{1}}]L^{-}[G^{-}_{\Delta_{2}}]\rangle=\frac{\Gamma(\Delta_{1}+\Delta_{2}-2)}{(iu_{21})^{\Delta_{1}+\Delta_{2}-2}}z_{21}^{\Delta_{1}}\bar{z}_{12}^{\Delta_{2}}. \label{2ptLightGraviton}
\end{align}
Similarly, the light-transformed three-point function of gravitons is,
\begin{align}
    \langle L^{+}[G^{+}_{\Delta_{1}}]L^{+}[G^{+}_{\Delta_{2}}]L^{-}[G^{-}_{\Delta_{3}}]\rangle& = (-1)^{\Delta_{1}+\Delta_{2}}\Gamma(\beta-2)z_{31}^{\Delta_{1}}z_{32}^{\Delta_{2}}\bar{z}_{12}^{\Delta_{3}+2}\mathrm{sgn}(\bar{z}_{21})\nn \\
    &\times \int_{\bar{z}_{1}}^{\bar{z}_{2}} d\bar{w}_{3}\frac{(\bar{z}_{2}-\bar{w}_{3})^{\Delta_{1}-2}(\bar{w}_{3}-\bar{z}_{1})^{\Delta_{2}-2}(\bar{w}_{3}-\bar{z}_{3})^{\Delta_{3}}}{[i(\bar{z}_{1}u_{32}+\bar{z}_{2}u_{13}+\bar{w}_{3}u_{21})]^{\beta-2}}. \label{3ptLightGraviton}
\end{align}
 We make the choice of $\mathrm{sgn}(\bar{z}_{21})=1$ for simplicity where it is understood that all the results that follow, in this section, will have an extra factor of $-1$ for the choice $\mathrm{sgn}(\bar{z}_{21})=-1$. We take the OPE limit $(u_{2},z_{2},\bar{z}_{2})\rightarrow (u_{3},z_{3},\bar{z}_{3})$ which we execute by performing an expansion of the above correlation function about $(u_{23},z_{23},\bar{z}_{23})=0$ \cite{Belavin:1984vu}. We start by taking the $u_{23}\rightarrow0$ limit \footnote{ One could, in principle, take a covariant point of view while bringing the points 1 and 2 close. However, $\mathcal{I}^+$ being a Carroll manifold, 
 spatial intervals are invariant under boosts, irrespective of the temporal interval between two points. Hence, the process of bringing two points close in the spatial sense can be done in a frame-independent manner. We adopted this procedure here, by considering temporal coincidence first, before probing possible singularity structure in OPE in spatial separation.} and obtain, in the leading order,
\begin{align}
   \langle L^{+}[G^{+}_{\Delta_{1}}]L^{+}[G^{+}_{\Delta_{2}}]L^{-}[G^{-}_{\Delta_{3}}]\rangle & \approx (-1)^{\Delta_{1}+\Delta_{2}}\frac{\Gamma(\beta-2)}{(iu_{13})^{\beta-2}}z_{31}^{\Delta_{1}}z_{32}^{\Delta_{2}}\bar{z}_{12}^{\Delta_{3}+2} \nn \\
   & \times \int_{\bar{z}_{1}}^{\bar{z}_{2}} d\bar{w}_{3} (\bar{z}_{2}-\bar{w}_{3})^{-\Delta_{2}-\Delta_{3}}(\bar{w}_{3}-\bar{z}_{1})^{\Delta_{2}-2}(\bar{w}_{3}-\bar{z}_{3})^{\Delta_{3}}. 
\end{align}
The integral evaluates to,
\begin{equation}
 \int_{\bar{z}_{1}}^{\bar{z}_{2}} d\bar{w}_{3} (\bar{z}_{2}-\bar{w}_{3})^{-\Delta_{2}-\Delta_{3}}(\bar{w}_{3}-\bar{z}_{1})^{\Delta_{2}-2}(\bar{w}_{3}-\bar{z}_{3})^{\Delta_{3}} = B(\Delta_{2}-1,1-\Delta_{2}-\Delta_{3})\frac{\bar{z}_{31}^{\Delta_{2}+\Delta_{3}-1}\bar{z}_{23}^{1-\Delta_{2}}}{\bar{z}_{21}^{1+\Delta_{3}}}
\end{equation}
where $B(x,y)$ is the Euler-Beta function. Using this result, we obtain,
\begin{align}
    \langle L^{+}[G^{+}_{\Delta_{1}}]L^{+}[G^{+}_{\Delta_{2}}]L^{-}[G^{-}_{\Delta_{3}}]\rangle & \approx -B(\Delta_{2}-1,1-\Delta_{2}-\Delta_{3})\bar{z}_{23}^{1-\Delta_{2}}z_{32}^{\Delta_{2}} \bar{z}_{12} \nn \\
    & \times \Bigg ( (-1)^\beta \Gamma(\beta-2) \frac{\bar{z}_{13}^{\Delta_{2}+\Delta_{3}-1}z_{31}^{\Delta_{1}}}{(iu_{13})^{\beta-2}}   \Bigg).
\end{align}
Now, $\bar{z}_{12}=\bar{z}_{13}-\bar{z}_{23}$, which gives us our leading and sub-leading term in the OPE as,
\begin{align}
    \langle L^{+}[G^{+}_{\Delta_{1}}]L^{+}[G^{+}_{\Delta_{2}}]&L^{-}[G^{-}_{\Delta_{3}}]\rangle  \approx -B(\Delta_{2}-1,1-\Delta_{2}-\Delta_{3})\bar{z}_{23}^{1-\Delta_{2}}z_{32}^{\Delta_{2}}\Bigg ( (-1)^\beta \Gamma(\beta-2) \frac{\bar{z}_{13}^{\Delta_{2}+\Delta_{3}}z_{31}^{\Delta_{1}}}{(iu_{13})^{\beta-2}}\Bigg) \nn \\
    & +B(\Delta_{2}-1,1-\Delta_{2}-\Delta_{3})\bar{z}_{23}^{2-\Delta_{2}}z_{32}^{\Delta_{2}}\times \Bigg ( (-1)^\beta \Gamma(\beta-2) \frac{\bar{z}_{13}^{\Delta_{2}+\Delta_{3}-1}z_{31}^{\Delta_{1}}}{(iu_{13})^{\beta-2}}\Bigg)
\end{align}
The first term in brackets is the two-point function of light-transformed graviton operators. Now, the sub-leading term can be expressed as, 
\begin{align}
    B(\Delta_{2}-1,1-\Delta_{2}-\Delta_{3})\bar{z}_{23}^{2-\Delta_{2}}z_{32}^{\Delta_{2}}\Bigg ( (-1)^\beta \Gamma(\beta-2) \frac{\bar{z}_{13}^{\Delta_{2}+\Delta_{3}-1}z_{31}^{\Delta_{1}}}{(iu_{13})^{\beta-2}}\Bigg)& = -\frac{B(\Delta_{2}-1,1-\Delta_{2}-\Delta_{3})}{\Delta_{2}+\Delta_{3}}\bar{z}_{23}^{2-\Delta_{2}}z_{32}^{\Delta_{2}}\nn \\
    & \times \Bigg ( (-1)^\beta \Gamma(\beta-2) \frac{\partial_{\bar{z}_{3}}\bar{z}_{13}^{\Delta_{2}+\Delta_{3}}z_{31}^{\Delta_{1}}}{(iu_{13})^{\beta-2}}\Bigg). 
\end{align}
Hence, in terms of the light-transformed two-point function of gravitons, the collinear limit takes the following form,
\begin{align}
   \langle L^{+}[G^{+}_{\Delta_{1}}]L^{+}[G^{+}_{\Delta_{2}}]&L^{-}[G^{-}_{\Delta_{3}}]\rangle \approx  -B(\Delta_{2}-1,1-\Delta_{2}-\Delta_{3})\bar{z}_{23}^{1-\Delta_{2}}z_{32}^{\Delta_{2}} \langle L^{+}[G^{+}_{\Delta_{1}}]L^{-}[G^{-}_{\Delta_{2}+\Delta_{3}}] \rangle \nn \\
   & -\frac{B(\Delta_{2}-1,1-\Delta_{2}-\Delta_{3})}{\Delta_{2}+\Delta_{3}}\bar{z}_{23}^{2-\Delta_{2}}z_{32}^{\Delta_{2}} \langle L^{+}[G^{+}_{\Delta_{1}}]\partial_{\bar{z}_{3}}L^{-}[G^{-}_{\Delta_{2}+\Delta_{3}}] \rangle 
\end{align}
Therefore, at this point, it becomes straightforward to read off the leading and sub-leading terms in the OPE as, 
\begin{align}
    L^{+}[G^{+}_{\Delta_{2}}]L^{-}[G^{-}_{\Delta_{3}}]&=-B(\Delta_{2}-1,1-\Delta_{2}-\Delta_{3})\bar{z}_{23}^{1-\Delta_{2}}z_{32}^{\Delta_{2}}L^{-}[G^{-}_{\Delta_{2}+\Delta_{3}}] \nn \\
    & -\frac{B(\Delta_{2}-1,1-\Delta_{2}-\Delta_{3})}{\Delta_{2}+\Delta_{3}}\bar{z}_{23}^{2-\Delta_{2}}z_{32}^{\Delta_{2}}\partial_{\bar{z}_{3}}L^{-}[G^{-}_{\Delta_{2}+\Delta_{3}}] .\label{CollinearGravitonOPE}
\end{align}
We must note that even though we have derived this OPE from the collinear limit of an anti-MHV amplitude, one could, do a similar calculation using the MHV correlator, $\langle L[G^{-}_{\Delta_{1}}]L[G^{-}_{\Delta_{2}}]L[G^{+}_{\Delta_{3}}]\rangle$ and derive $L^{-}[G^{-}_{\Delta_{2}}]L^{+}[G^{+}_{\Delta_{3}}]$\footnote{We thank Sruthi Narayanan for bringing this point to our notice}. 
In the next section, through an explicit symmetry check, we will assert the significance of keeping the sub-leading term of the above operator algebra.

\subsection{Translation symmetry analysis of graviton OPE} \label{gravitonSymmetry}
In this section, we probe the importance of the subleading term in the OPE \eqref{CollinearGravitonOPE}. We start by noting that, trivially, the leading term contribution to this OPE is invariant in null-time translation $M_{00} \sim \partial_u$. Nevertheless, checking the invariance of the OPE, \eqref{CollinearGravitonOPE}, under $\delta_{M_{01}}$, $\delta_{M_{10}}$ and $\delta_{M_{11}}$ are non-trivial owing to the appearance of linear powers of $z$ and $\bar{z}$ in their expressions. In \cite{Banerjee:2024hvb}, we pursued this non-triviality and observed that on the space of light transformed operators, the actions of supertranslations, including the time-translation generators, can be expressed in terms of vector fields tangential to the celestial sphere. We restate the results here,
\begin{align}
    &\delta_{M_{00}}L^{+}[\phi_{a,\mathfrak{h}}]= -\frac{i\epsilon}{1-2h} \partial_{z}L^{+}[\phi_{a,\mathfrak{h}+\frac{1}{2}}] \label{L+M00} \\
    &\delta_{M_{00}}L^{-}[\phi_{a,\mathfrak{h}}]= -\frac{i\epsilon}{1-2\bar{h}} \partial_{\bar{z}}L^{-}[\phi_{a,\mathfrak{h}+\frac{1}{2}}] \label{L-M00}\\
    & \delta_{M_{10}}L^{+}[\phi_{a,\mathfrak{h}}]=-i\epsilon \Big( \frac{z}{1-2h}\partial_{z}+1 \Big)L^{+}[\phi_{a,\mathfrak{h}+\frac{1}{2}}] \label{L+M10} \\
    & \delta_{M_{10}}L^{-}[\phi_{a,\mathfrak{h}}]=-\frac{i\epsilon z}{1-2\bar{h}}\partial_{\bar{z}}L^{-}[\phi_{a,\mathfrak{h}+\frac{1}{2}}] \label{L-M10} \\
& \delta_{M_{01}}L^{+}[\phi_{a,\mathfrak{h}}]=-\frac{i\epsilon \bar{z}}{1-2h}\partial_{z}L^{+}[\phi_{a,\mathfrak{h}+\frac{1}{2}}] \label{L+M01} \\ 
    & \delta_{M_{01}}L^{-}[\phi_{a,\mathfrak{h}}]=-i\epsilon \Big( \frac{\bar{z}}{1-2\bar{h}}\partial_{\bar{z}}+1 \Big)L^{-}[\phi_{a,\mathfrak{h}+\frac{1}{2}}] \label{L-M01}\\
    & \delta_{M_{11}}L^{+}[\phi_{a,\mathfrak{h}}]=-i\epsilon \bar{z}\Big(\frac{z}{1-2h}\partial_{z}+ 1\Big)L^{+}[\phi_{a,\mathfrak{h}+\frac{1}{2}}] \label{L+M11}\\
    & \delta_{M_{11}}L^{-}[\phi_{a,\mathfrak{h}}]=-i\epsilon z \Big(\frac{\bar{z}}{1-2\bar{h}}\partial_{\bar{z}}+1\Big)L^{-}[\phi_{a,\mathfrak{h}+\frac{1}{2}}]  \label{L-M11}
\end{align}
Due to (a) the presence of a spatial derivative and (b) $M_{00}$ not acting diagonally on the basis of conformal weights, by shifting $\mathfrak{h} \rightarrow \mathfrak{h}+\frac{1}{2}$, there are some potential non-trivialities involved in checking the symmetries of \eqref{CollinearGravitonOPE}.

To proceed further, we evaluate the transformation $\delta_{M_{00}}$ on the left-hand side of \eqref{CollinearGravitonOPE} 
using the form of \eqref{L+M00} and \eqref{L-M00} in the context of gravitons with proper light transforms:
\begin{equation}
    \delta_{M_{00}}L^{+}[G^{+}_{\Delta}]=\frac{i \epsilon \partial_{z}}{\Delta+1}L^{+}[G^{+}_{\Delta+1}] \qquad \text{and } \qquad \delta_{M_{00}}L^{-}[G^{-}_{\Delta}]=\frac{i \epsilon \partial_{\bar{z}}}{\Delta+1}L^{+}[G^{+}_{\Delta+1}].
\end{equation}
where we choose $-\epsilon_{G^{+}_{\Delta_{2}}}=\epsilon_{G^{-}_{\Delta_{3}}}=\epsilon_{G^{-}_{\Delta_{2}+\Delta_{3}}}=  1$. This gives us,
\begin{align}
  &-B(\Delta_{2},-\Delta_{2}-\Delta_{3})\bar{z}_{23}^{-\Delta_{2}}\big(\frac{-i\partial_{z_{2}}}{\Delta_{2}+1}   \big)z_{32}^{\Delta_{2}+1}L^{-}[G^{-}_{\Delta_{2}+\Delta_{3}+1}]\nn \\
  &-\frac{B(\Delta_{2},-\Delta_{2}-\Delta_{3})}{\Delta_{2}+\Delta_{3}+1}\bar{z}_{23}^{1-\Delta_{2}}\big(\frac{-i\partial_{z_{2}}}{\Delta_{2}+1}   \big)z_{32}^{\Delta_{2}+1}\partial_{\bar{z}_{3}}L^{-}[G^{-}_{\Delta_{2}+\Delta_{3}+1}]  \nn \\
  &-B(\Delta_{2}-1,-\Delta_{2}-\Delta_{3})z_{32}^{\Delta_{2}}\big( \frac{i \partial_{\bar{z}_{3}}}{\Delta_{3}+1}\big)\big(\bar{z}_{23}^{1-\Delta_{2}}L^{-}[G^{-}_{\Delta_{2}+\Delta_{3}+1}] \big) \nn \\
  &-\frac{B(\Delta_{2}-1,-\Delta_{2}-\Delta_{3})}{\Delta_{2}+\Delta_{3}}z_{32}^{\Delta_{2}}\big( \frac{i \partial_{\bar{z}_{3}}}{\Delta_{3}+1}\big)\big(\bar{z}_{23}^{2-\Delta_{2}}\partial_{\bar{z}_{3}}L^{-}[G^{-}_{\Delta_{2}+\Delta_{3}+1}]\big)
\end{align}
Collecting similar powers of $z_{32}$ and $\bar{z}_{23}$ we arrive at,
\begin{align}
    &i\Bigg[ -B(\Delta_{2},-\Delta_{2}-\Delta_{3})+B(\Delta_{2}-1,-\Delta_{2}-\Delta_{3})\frac{1-\Delta_{1}}{\Delta_{2}+1}\Bigg]\bar{z}_{23}^{-\Delta_{2}}z_{32}^{\Delta_{2}}L^{-}[G^{-}_{\Delta_{2}+\Delta_{3}+1}] \nn \\
    &+i \Bigg[-\color{magenta}\frac{B(\Delta_{2},-\Delta_{2}-\Delta_{3})}{\Delta_{2}+\Delta_{3}+1}\color{black}-\frac{B(\Delta_{2}-1,-\Delta_{2}-\Delta_{3})}{\Delta_{3}+1} \nn \\
    & +\color{magenta}B(\Delta_{2}-1,-\Delta_{2}-\Delta_{3})\frac{2-\Delta_{2}}{\Delta_{3}+1}\color{black}\Bigg]\bar{z}_{23}^{1-\Delta_{2}}z_{32}^{\Delta_{2}}\partial_{\bar{z}_{3}}L^{-}[G^{-}_{\Delta_{2}+\Delta_{3}+1}]\nn \\
    &-i \frac{B(\Delta_{2}-1,-\Delta_{2}-\Delta_{3})}{(\Delta_{2}+\Delta_{3}+1)(\Delta_{3}+1)}\bar{z}_{23}^{2-\Delta_{2}}z_{32}^{\Delta_{2}}\partial^{2}_{\bar{z}_{3}}L^{-}[G^{-}_{\Delta_{2}+\Delta_{3}+1}]. \label{step1}
\end{align}
Using the properties of the Euler-Beta function we get:
\begin{align}
   & -B(\Delta_{2},-\Delta_{2}-\Delta_{3})+B(\Delta_{2}-1,-\Delta_{2}-\Delta_{3})\frac{1-\Delta_{1}}{\Delta_{2}+1}=0 \nn \\
   & -\color{magenta}\frac{B(\Delta_{2},-\Delta_{2}-\Delta_{3})}{\Delta_{2}+\Delta_{3}+1}\color{black}-\frac{B(\Delta_{2}-1,-\Delta_{2}-\Delta_{3})}{\Delta_{3}+1}+\color{magenta}\frac{B(\Delta_{2}-1,-\Delta_{2}-\Delta_{3})(2-\Delta_{2})}{\Delta_{3}+1}\color{black}=-\frac{B(\Delta_{2}-1,1-\Delta_{2}-\Delta_{3})}{\Delta_{2}+\Delta_{3}+1} 
\end{align}
using this we further simplify the expression in \eqref{step1} as, 
\begin{align}
   &-i\frac{B(\Delta_{2}-1,1-\Delta_{2}-\Delta_{3})}{\Delta_{2}+\Delta_{3}+1} \bar{z}_{23}^{1-\Delta_{2}}z_{32}^{\Delta_{2}}\partial_{\bar{z}_{3}}L^{-}[G^{-}_{\Delta_{2}+\Delta_{3}+1}]  \nn \\
   &-i \frac{B(\Delta_{2}-1,1-\Delta_{2}-\Delta_{3})}{(\Delta_{2}+\Delta_{3}+1)(\Delta_{2}+\Delta_{3})}\bar{z}_{23}^{2-\Delta_{2}}z_{32}^{\Delta_{2}}\partial^{2}_{\bar{z}_{3}}L^{-}[G^{-}_{\Delta_{2}+\Delta_{3}+1}].\label{M00GravitonOPELHS}
\end{align}
On the other hand, applying $\delta_{M_{00}}$ on the right-hand side of \eqref{CollinearGravitonOPE}, we get the following: 
\begin{align}
    &-i\frac{B(\Delta_{2}-1,1-\Delta_{2}-\Delta_{3})}{\Delta_{2}+\Delta_{3}+1} \bar{z}_{23}^{1-\Delta_{2}}z_{32}^{\Delta_{2}}\partial_{\bar{z}_{3}}L^{-}[G^{-}_{\Delta_{2}+\Delta_{3}+1}] \nn\\ 
    &-i \frac{B(\Delta_{2}-1,-\Delta_{2}-\Delta_{3})}{(\Delta_{2}+\Delta_{3})^{2}}\bar{z}_{23}^{2-\Delta_{2}}z_{32}^{\Delta_{2}}\partial^{2}_{\bar{z}_{3}}L^{-}[G^{-}_{\Delta_{2}+\Delta_{3}+1}]. \label{M00GravitonOPERHS}
\end{align}
Comparing the leading term in \eqref{M00GravitonOPELHS} and \eqref{M00GravitonOPERHS}, we see that the OPE is indeed symmetric under the transformation $\delta_{M_{00}}$ in the leading order. One must note that without the terms in magenta, which happen to be the contributions from the sub-leading order, the action of $\delta_{M_{00}}$ on the left-hand side and the right-hand side of \eqref{CollinearGravitonOPE} would not match in the leading order. Also, the same conclusion is drawn while checking the symmetry properties of the above OPE for $\delta_{M_{01}}, \delta_{M_{10}}$ and $\delta_{M_{11}}$. As is evident from the above equation the sub-leading terms do not match on both sides. This is because we have not considered the contribution of sub$^{2}$-leading terms in the OPE. If one keeps the sub$^{2}$-leading term in the OPE, then there would be terms which would have contributed to the last line of \eqref{step1} such that they would have added up to give the second term appearing in \eqref{M00GravitonOPERHS}. In general, while checking the translation symmetry of the graviton OPE, every order receives a contribution from the next order.
\section{General structure of the operator product expansion} \label{General structure of the operator product expansion}
In this section, we show that in the light-transformed basis, one can find a class of OPEs from a purely intrinsic approach. ie. using only the global part of supertranslations. We start with the usual CFT ansatz,
\begin{align}
    L^{+}[\phi_{\mathfrak{h}_{1}}]L^{-}[\phi_{\mathfrak{h}_{2}}]&=C(\mathfrak{h}_{1},\mathfrak{h}_{2})z_{12}^{\tilde{h}_{p}-\tilde{h}_{1}-\tilde{h}_{2}}\bar{z}_{12}^{\tilde{\bar{h}}_{p}-\tilde{\bar{h}}_{1}-\tilde{\bar{h}}_{2}}L^{-}[\phi_{\mathfrak{h}_{p}}]\nn \\ 
    &+ D(\mathfrak{h}_{1},\mathfrak{h}_{2})z_{12}^{\tilde{h}_{p}-\tilde{h}_{1}-\tilde{h}_{2}}\bar{z}_{12}^{\tilde{\bar{h}}_{p}-\tilde{\bar{h}}_{1}-\tilde{\bar{h}}_{2}+1}\partial_{\bar{z}_{2}}L^{-}[\phi_{\mathfrak{h}_{p}}] \label{OPELightAnsatz}
\end{align}
where we have chosen to keep the sub-leading term, motivated by our calculations from the previous section. 
Also, one must note that owing to the fact that these operators are light-transformed Carroll primaries, we expect terms with powers of $u_{12}$ on the right-hand side. However, we do not consider such terms here because, as we mentioned before (eg, see the footnote on page 4), one can consider OPEs in time and space separately on Carrollian manifolds, owing to the fact that the spatial intervals on their own are frame invariant. We will see that keeping the sub-leading term not only gives us the correct expression for the scaling dimension of the operator appearing on the right-hand side of \eqref{GeneralOPEAnsatz} but also aids in calculating the OPE coefficient corresponding to the leading term, i.e $C(\mathfrak{h}_{1},\mathfrak{h}_{2})$, once the spin of all the operators are specified. We will explicitly check the robustness of this method by computing the leading term in the OPE \eqref{CollinearGravitonOPE} without using any bulk data in section \ref{IntrinsicGraviton}.

In the above equation, the twiddled weights are related to the un-twiddled ones by,
\begin{align}
   & \tilde{h}_{1}=1-h_{1} \qquad \text{and} \qquad \tilde{\bar{h}}_{1}=\bar{h}_{1} \\
   & \tilde{h}_{2}=h_{2} \qquad \text{and} \qquad \tilde{\bar{h}}_{2}=1-\bar{h}_{2} \\
   & \tilde{h}_{p}=h_{p} \qquad \text{and} \qquad \tilde{\bar{h}}_{p}=1-\bar{h}_{p}. 
\end{align}
In terms of the untwiddled weights, the OPE takes the form,
\begin{align}
     L^{+}[\phi_{\mathfrak{h}_{1}}]L^{-}[\phi_{\mathfrak{h}_{2}}]&=C(\mathfrak{h}_{1},\mathfrak{h}_{2})z_{12}^{h_{p}+h_{1}-h_{2}-1}\bar{z}_{12}^{\bar{h}_{2}-\bar{h}_{p}-\bar{h}_{1}}L^{-}[\phi_{\mathfrak{h}_{p}}]\nn \\ 
    &+ D(\mathfrak{h}_{1},\mathfrak{h}_{2})z_{12}^{h_{p}+h_{1}-h_{2}-1}\bar{z}_{12}^{\bar{h}_{2}-\bar{h}_{p}-\bar{h}_{1}+1}\partial_{\bar{z}_{2}}L^{-}[\phi_{\mathfrak{h}_{p}}] \label{GeneralOPEAnsatz}
\end{align}
where each operator has an associated kinematic index $\epsilon_{i}$ which labels whether the i'th particle is incoming or outgoing. For our calculations we choose $\epsilon_{1}=-\epsilon_{2}=-\epsilon_{p}=-\epsilon$ where $\epsilon=\pm1$. We first attempt to fix the scaling dimension of the operator appearing in the leading order. This will, of course, fix the scaling dimension of all the other operators appearing in the sub-leading terms.
\subsection{Fixing the scaling dimension} \label{generalOPEScaling}
The action of $\delta_{M_{00}}$ on the OPE is given by,
\begin{align}
    \delta_{M_{00}}L^{+}[\phi_{\mathfrak{h}_{1}}]L^{-}[\phi_{\mathfrak{h}_{2}}]+ L^{+}[\phi_{\mathfrak{h}_{1}}]\delta_{M_{00}}L^{-}[\phi_{\mathfrak{h}_{2}}]&=C(\mathfrak{h}_{1},\mathfrak{h}_{2})z_{12}^{h_{p}+h_{1}-h_{2}-1}\bar{z}_{12}^{\bar{h}_{2}-\bar{h}_{p}-\bar{h}_{1}}\delta_{M_{00}}L^{-}[\phi_{\mathfrak{h}_{p}}]\nn \\ 
    &+ D(\mathfrak{h}_{1},\mathfrak{h}_{2})z_{12}^{h_{p}+h_{1}-h_{2}-1}\bar{z}_{12}^{\bar{h}_{2}-\bar{h}_{p}-\bar{h}_{1}+1}\partial_{\bar{z}_{2}}\delta_{M_{00}}L^{-}[\phi_{\mathfrak{h}_{p}}].
\end{align}
We recall that the action of $\delta_{M_{00}}$ is given by,
\begin{align}
    \delta_{M_{00}}L^{+}[\phi_{\mathfrak{h}}]=-i\epsilon\frac{\partial_{z}}{1-2h}L^{+}[\phi_{\mathfrak{h}+\frac{1}{2}}] \qquad \text{and} \qquad \delta_{M_{00}}L^{-}[\phi_{\mathfrak{h}}]=-i\epsilon\frac{\partial_{\bar{z}}}{1-2\bar{h}}L^{}[\phi_{\mathfrak{h}+\frac{1}{2}}]. \label{M00ActionLight}
\end{align}
Using this, we get,
\begin{align}
  &i\epsilon C(\mathfrak{h}_{1}+\frac{1}{2},\mathfrak{h}_{2})\frac{\partial_{z_{1}}}{1-2h_{1}}z_{12}^{h_{p}+h_{1}-h_{2}}\bar{z}_{12}^{\bar{h}_{2}-\bar{h}_{p}-\bar{h}_{1}-1}L^{-}[\phi_{\mathfrak{h}_{p}+\frac{1}{2}}] \nn \\
  &-i\epsilon C(\mathfrak{h}_{1},\mathfrak{h}_{2}+\frac{1}{2})z_{12}^{h_{p}+h_{1}-h_{2}-1}\frac{\partial_{\bar{z}_{2}}}{1-2\bar{h}_{2}}\Big(\bar{z}_{12}^{\bar{h}_{2}-\bar{h}_{p}-\bar{h}_{1}}L^{-}[\phi_{\mathfrak{h}_{p}+\frac{1}{2}}]\Big) \nn \\
&+i\epsilon D(\mathfrak{h}_{1}+\frac{1}{2},\mathfrak{h}_{2})\frac{\partial_{z_{1}}}{1-2h_{1}}z_{12}^{h_{p}+h_{1}-h_{2}}\bar{z}_{12}^{\bar{h}_{2}-\bar{h}_{p}-\bar{h}_{1}}\partial_{\bar{z}_{2}}L^{-}[\phi_{\mathfrak{h}_{p}+\frac{1}{2}}]\nn \\
&-i\epsilon D(\mathfrak{h}_{1},\mathfrak{h}_{2}+\frac{1}{2})z_{12}^{h_{p}+h_{1}-h_{2}-1}\frac{\partial_{\bar{z}_{2}}}{1-2\bar{h}_{2}}\Big(\bar{z}_{12}^{\bar{h}_{2}-\bar{h}_{p}-\bar{h}_{1}+1}\partial_{\bar{z}_{2}}L^{-}[\phi_{\mathfrak{h}_{p}+\frac{1}{2}}]\Big) \nn\\
& =-i\epsilon C(\mathfrak{h}_{1},\mathfrak{h}_{2})z_{12}^{h_{p}+h_{1}-h_{2}-1}\bar{z}_{12}^{\bar{h}_{2}-\bar{h}_{p}-\bar{h}_{1}}\Big(\frac{\partial_{\bar{z}_{2}}}{1-2\bar{h}_{p}}\Big)L^{-}[\phi_{\mathfrak{h}_{p}+\frac{1}{2}}] \nn \\
&-i\epsilon D(\mathfrak{h}_{1},\mathfrak{h}_{2})z_{12}^{h_{p}+h_{1}-h_{2}-1}\bar{z}_{12}^{\bar{h}_{2}-\bar{h}_{p}-\bar{h}_{1}+1}\partial_{\bar{z}_{2}}\Big(\frac{\partial_{\bar{z}_{2}}}{1-2\bar{h}_{p}}\Big)L^{-}[\phi_{\mathfrak{h}_{p}+\frac{1}{2}}]
\end{align}
from which we collect similar powers of $z_{12}$ and $\bar{z}_{12}$ and get,
\begin{align}
&i\epsilon \Big[\frac{C(\mathfrak{h}_{1}+\frac{1}{2},\mathfrak{h}_{2})}{1-2h_{1}}(h_{p}+h_{1}-h_{2})+\frac{C(\mathfrak{h}_{1},\mathfrak{h}_{2}+\frac{1}{2})}{1-2\bar{h}_{2}}(\bar{h}_{2}-\bar{h}_{p}-\bar{h}_{1})\Big]z_{12}^{h_{p}+h_{1}-h_{2}-1}\bar{z}_{12}^{\bar{h}_{2}-\bar{h}_{p}-\bar{h}_{1}-1}L^{-}[\phi_{\mathfrak{h}_{p}+\frac{1}{2}}]  \nn \\
& +i\epsilon \Big[\frac{D(\mathfrak{h}_{1}+\frac{1}{2},\mathfrak{h}_{2})}{1-2h_{1}}(h_{p}+h_{1}-h_{2})-\frac{C(\mathfrak{h}_{1},\mathfrak{h}_{2}+\frac{1}{2})}{1-2\bar{h}_{2}}\nn \\
&+\frac{D(\mathfrak{h}_{1},\mathfrak{h}_{2}+\frac{1}{2})}{1-2\bar{h}_{2}}(\bar{h}_{2}-\bar{h}_{p}-\bar{h}_{1}+1) \Big]z_{12}^{h_{p}+h_{1}-h_{2}-1}\bar{z}_{12}^{\bar{h}_{2}-\bar{h}_{p}-\bar{h}_{1}}\partial_{\bar{z}_{2}}L^{-}[\phi_{\mathfrak{h}_{p}+\frac{1}{2}}]\nn \\
&- i\epsilon \frac{D(\mathfrak{h}_{1},\mathfrak{h}_{2}+\frac{1}{2})}{1-2\bar{h}_{2}}z_{12}^{h_{p}+h_{1}-h_{2}-1}\bar{z}_{12}^{\bar{h}_{2}-\bar{h}_{p}-\bar{h}_{1}}\partial^{2}_{\bar{z}_{2}}L^{-}[\phi_{\mathfrak{h}_{p}+\frac{1}{2}}] \nn\\
& =-i\epsilon C(\mathfrak{h}_{1},\mathfrak{h}_{2})z_{12}^{h_{p}+h_{1}-h_{2}-1}\bar{z}_{12}^{\bar{h}_{2}-\bar{h}_{p}-\bar{h}_{1}}\Big(\frac{\partial_{\bar{z}_{2}}}{1-2\bar{h}_{p}}\Big)L^{-}[\phi_{\mathfrak{h}_{p}+\frac{1}{2}}] \nn \\
&-i\epsilon D(\mathfrak{h}_{1},\mathfrak{h}_{2})z_{12}^{h_{p}+h_{1}-h_{2}-1}\bar{z}_{12}^{\bar{h}_{2}-\bar{h}_{p}-\bar{h}_{1}+1}\partial_{\bar{z}_{2}}\Big(\frac{\partial_{\bar{z}_{2}}}{1-2\bar{h}_{p}}\Big)L^{-}[\phi_{\mathfrak{h}_{p}+\frac{1}{2}}].
\end{align}
Comparing powers of $z_{12}$ and $\bar{z}_{12}$ in the leading order, we get,
\begin{align}
    &\frac{C(\mathfrak{h}_{1}+\frac{1}{2},\mathfrak{h}_{2})}{1-2h_{1}}(h_{p}+h_{1}-h_{2})+\frac{C(\mathfrak{h}_{1},\mathfrak{h}_{2}+\frac{1}{2})}{1-2\bar{h}_{2}}(\bar{h}_{2}-\bar{h}_{p}-\bar{h}_{1}) =0 \label{generalRecursion1} \\
    &\frac{D(\mathfrak{h}_{1}+\frac{1}{2},\mathfrak{h}_{2})}{1-2h_{1}}(h_{p}+h_{1}-h_{2})-\frac{C(\mathfrak{h}_{1},\mathfrak{h}_{2}+\frac{1}{2})}{1-2\bar{h}_{2}}+\frac{D(\mathfrak{h}_{1},\mathfrak{h}_{2}+\frac{1}{2})}{1-2\bar{h}_{2}}(\bar{h}_{2}-\bar{h}_{p}-\bar{h}_{1}+1)=-\frac{C(\mathfrak{h}_{1},\mathfrak{h}_{2})}{1-2\bar{h}_{p}} \label{generalRecursion2}
\end{align}
Further, we use the $\delta_{M_{01}}$ symmetry in the same spirit as shown and get the following relations, 
\begin{align}
   &C(\mathfrak{h}_{1},\mathfrak{h}_{2}+\frac{1}{2})-\frac{C(\mathfrak{h}_{1}+\frac{1}{2},\mathfrak{h}_{2})}{1-2h_{1}}(h_{p}+h_{1}-h_{2})=C(\mathfrak{h}_{1},\mathfrak{h}_{2}) \label{generalRecursion3} \\
   &D(\mathfrak{h}_{1},\mathfrak{h}_{2}+\frac{1}{2})-\frac{D(\mathfrak{h}_{1}+\frac{1}{2},\mathfrak{h}_{2})}{1-2h_{1}}(h_{p}+h_{1}-h_{2})=D(\mathfrak{h}_{1},\mathfrak{h}_{2})\bigg( \frac{2-2\bar{h}_{p}}{1-2\bar{h}_{p}}\bigg).\label{generalRecursion4}
\end{align}
Similarly, from $\delta_{M_{10}}$ we get, 
\begin{align}
    &\frac{C(\mathfrak{h}+\frac{1}{2},\mathfrak{h}_{2})}{1-2h_{1}}(h_{p}+h_{1}-h_{2})+C(\mathfrak{h}_{1}+\frac{1}{2},\mathfrak{h}_{2})=0 \label{generalRecursion5} \\
    &\frac{D(\mathfrak{h}+\frac{1}{2},\mathfrak{h}_{2})}{1-2h_{1}}(h_{p}+h_{1}-h_{2})+D(\mathfrak{h}_{1}+\frac{1}{2},\mathfrak{h}_{2})=0 \label{generalRecursion6}.
\end{align}
Owing to the cumbersome nature of the expressions appearing while performing these calculations, we delegate them to Appendix \ref{translationSymmetry}. Now, by demanding that $C(\mathfrak{h}+\frac{1}{2},\mathfrak{h}_{2})\neq0$ we immediately conclude, from \eqref{generalRecursion5}, that, 
\begin{equation}
    \boxed{h_{p}=h_{1}+h_{2}-1} \label{generalScaling}
\end{equation}
This is one of the central results of this paper. The most remarkable thing about working with the light-transformed basis is that we have not referred to any bulk theory in deriving the above result. The authors have not come across any such basis in the literature before. We know that in the context of flat holography, the dimension of the operator appearing in the leading term of the OPE is determined from the bulk interaction vertex \cite{Pate:2019lpp}. What we have shown is that the formula \eqref{generalScaling} predicts the existence of these vertices from an intrinsic perspective. What is intriguing is that, as we show in the following section, along with the well-known results, this formula also predicts the existence of certain `exotic' vertices; however, finding bulk theories corresponding to those vertices is beyond the scope of this work.

\subsection{Fixing the leading OPE coefficient} \label{Fixing the leading OPE coefficient}
Given equations \eqref{generalRecursion1} to \eqref{generalRecursion4} and the solution to $h_{p}$ as given in \eqref{generalScaling}, one can, in principle, fix the coefficient of the leading term provided we choose a value for the spin $s_{p}$. To see that, we first simplify the aforementioned expression as,
\begin{align}
    &-C(\mathfrak{h}_{1}+\frac{1}{2},\mathfrak{h}_{2})+\frac{C(\mathfrak{h}_{1},\mathfrak{h}_{2}+\frac{1}{2})}{1-2\bar{h}_{2}}(\bar{h}_{2}-\bar{h}_{p}-\bar{h}_{1}) =0 \label{generalRecursion1.1} \\
    &-D(\mathfrak{h}_{1}+\frac{1}{2},\mathfrak{h}_{2})-\frac{C(\mathfrak{h}_{1},\mathfrak{h}_{2}+\frac{1}{2})}{1-2\bar{h}_{2}}+\frac{D(\mathfrak{h}_{1},\mathfrak{h}_{2}+\frac{1}{2})}{1-2\bar{h}_{2}}(\bar{h}_{2}-\bar{h}_{p}-\bar{h}_{1}+1)=-\frac{C(\mathfrak{h}_{1},\mathfrak{h}_{2})}{1-2\bar{h}_{p}} \label{generalRecursion1.2}\\
    &C(\mathfrak{h}_{1},\mathfrak{h}_{2}+\frac{1}{2})+C(\mathfrak{h}_{1}+\frac{1}{2},\mathfrak{h}_{2})=C(\mathfrak{h}_{1},\mathfrak{h}_{2})\label{generalRecursion1.3} \\
   &D(\mathfrak{h}_{1},\mathfrak{h}_{2}+\frac{1}{2})+D(\mathfrak{h}_{1}+\frac{1}{2},\mathfrak{h}_{2})=D(\mathfrak{h}_{1},\mathfrak{h}_{2})\bigg( \frac{2-2\bar{h}_{p}}{1-2\bar{h}_{p}}\bigg).\label{generalRecursion1.4}
\end{align}
Formally speaking, equations \eqref{generalRecursion1.1} and \eqref{generalRecursion1.3} are sufficient to solve for $C(\mathfrak{h}_{1},\mathfrak{h}_{2})$ once we have made a choice for the spin $s_{p}$ and fixed the scaling dimension $\Delta_{p}$ of the operator $\phi_{\mathfrak{h}_{p}}$ using \eqref{generalScaling}. In the next few sections, we work out explicitly known examples of OPEs of various field theories. When writing the OPE coefficients we change our notation to the more familiar notation of \cite{Pate:2019lpp}, wherein it is understood that the OPE coefficients are functions of $\Delta_{1}$ and $\Delta_{2}$, so from now onwards 
$$C(\mathfrak{h}_{1},\mathfrak{h}_{2}) \equiv C(\Delta_{1},\Delta_{2}).$$
\subsubsection{Gravitons} \label{IntrinsicGraviton}
Let us consider an OPE of the form,
\begin{align}
     L^{+}[G^{+}_{\Delta_{1}}]L^{-}[G^{-}_{\Delta_{2}}]&=C(\Delta_{1},\Delta_{2})z_{12}^{h_{p}+h_{1}-h_{2}-1}\bar{z}_{12}^{\bar{h}_{2}-\bar{h}_{p}-\bar{h}_{1}}L^{-}[G^{-}_{\Delta_{p}}]\nn \\ 
    &+ D(\Delta_{1},\Delta_{2})z_{12}^{h_{p}+h_{1}-h_{2}-1}\bar{z}_{12}^{\bar{h}_{2}-\bar{h}_{p}-\bar{h}_{1}+1}\partial_{\bar{z}_{2}}L^{-}[G^{-}_{\Delta_{p}}]. \label{GeneralGravitonOPE}
\end{align}
where $G^{\pm}_{\Delta}$ represents a graviton of scaling dimension $\Delta$ and helicity $\pm1$. Fixing the $4D$ helicity of the graviton amounts to fixing its $2D$ spin\footnote{Following this convention the 2D spin is identified with the 4D helicity, so for the case of gravitons $h_{p}=\frac{\Delta_{p}-2}{2}$, $h_{1}=\frac{\Delta_{1}+2}{2}$ and $h_{2}=\frac{\Delta_{2}-2}{2}$} \cite{Pasterski:2017ylz}. The formula \eqref{generalScaling} tells us that,
\begin{align}
  \Delta_{p}=\Delta_{1}+\Delta_{2} \label{gravitonScaling}
\end{align}
which matches with \cite{Banerjee:2020kaa}. This same result was obtained by analysing the mass dimension of the bulk three-point vertex in momentum space\cite{Pate:2019lpp}. Albeit the calculations are done in a different basis, what equation \eqref{gravitonScaling} tells us is that we can now infer physical information about the bulk from a boundary theory written in the light-transformed basis. Equations \eqref{generalRecursion1.1} and \eqref{generalRecursion1.3} are,
\begin{align}
    &C(\Delta_{1}+1,\Delta_{2})+\frac{C(\Delta_{1},\Delta_{2}+1)}{\Delta_{2}+1}(1-\Delta_{1})=0 \label{gravitonRecursion1} \\
    &C(\Delta_{1}+1,\Delta_{2})+C(\Delta_{1},\Delta_{2}+1)=C(\Delta_{1},\Delta_{2}) \label{gravitonRecursion2}.
\end{align}
These equations can be re-packaged as,
\begin{align}
    &C(\Delta_{1},\Delta_{2})(1+\Delta_{2})=(\Delta_{1}+\Delta_{2})C(\Delta_{1},\Delta_{2}+1) \label{gravitonrecursion3}\\
    &C(\Delta_{1},\Delta_{2})(\Delta_{1}-1)=(\Delta_{1}+\Delta_{2})C(\Delta_{1}+1,\Delta_{2}) \label{gravitonrecursion4}
\end{align}
and are easily solved if we make the following change of variables 
\begin{equation}
    (-1)^{\Delta_{1}}\tilde{C}(\Delta_{1},\Delta_{2})=C(\Delta_{1},\Delta_{2}) \label{newVariable}
\end{equation}
which implies,
\begin{align}
    &\tilde{C}(\Delta_{1},\Delta_{2})(1+\Delta_{2})=(\Delta_{1}+\Delta_{2})\tilde{C}(\Delta_{1},\Delta_{2}+1) \label{gravitonrecursion5}\\
    &\tilde{C}(\Delta_{1},\Delta_{2})(\Delta_{1}-1)=-(\Delta_{1}+\Delta_{2})\tilde{C}(\Delta_{1}+1,\Delta_{2}) \label{gravitonrecursion6}.
\end{align}
These equations are easily solved using Weiland's theorem \cite{db6dc5fa-41f3-3c6a-acf0-210f0aa62d5a}, and we have,
\begin{align}
   & \tilde{C}(\Delta_{1},\Delta_{2})=\mathcal{N}_{grav}B(\Delta_{1}-1,1-\Delta_{1}-,\Delta_{2})\nn \\
   &\implies \boxed{C(\Delta_{1},\Delta_{2})=\mathcal{N}_{grav}(-1)^{\Delta_{1}}B(\Delta_{1}-1,1-\Delta_{1}-,\Delta_{2})} \label{gravitonOPECoefficient}.
\end{align}
where $\mathcal{N}_{grav}$ is a numerical constant, independent of $\Delta_{1}$ and $\Delta_{2}$ \footnote{$\mathcal{N}_{grav}$ can be fixed, however by imposing a metric on the module of light-transformed operators}. So, using \eqref{gravitonOPECoefficient} and \eqref{gravitonScaling} we find that the OPE leading order term in \eqref{GeneralGravitonOPE} is, 
\begin{align}
     L^{+}[G^{+}_{\Delta_{1}}]L^{-}[G^{-}_{\Delta_{2}}]=\mathcal{N}_{grav}(-1)^{\Delta_{1}}B(\Delta_{1}-1,1-\Delta_{1}-,\Delta_{2})z_{12}^{\Delta_{1}}\bar{z}_{12}^{1-\Delta_{1}}L^{-}[G^{-}_{\Delta_{1}+\Delta_{2}}]. \label{IntrinsicGravitonOPE}
\end{align}
Up to a numerical constant, this is exactly the OPE \eqref{CollinearGravitonOPE} which was obtained in the previous section by studying the collinear limit of the Carrollian graviton correlator in the light-transformed basis. 
\subsubsection{Gluons}
In our general ansatz, we choose $\phi_{\mathfrak{h}_{1}}=O^{+}_{\Delta_{1}}$ and $\phi_{\mathfrak{h}_{2}}=O^{-}_{\Delta_{2}}$ where we use $O^{\pm}_{\Delta}$ to represent a gluon of scaling dimension $\Delta$ and helicity $\pm1$. The OPE we consider is of the form,
\begin{align}
     L^{+}[O^{+}_{\Delta_{1}}]L^{-}[O^{-}_{\Delta_{2}}]&=E(\Delta_{1},\Delta_{2})z_{12}^{h_{p}+h_{1}-h_{2}-1}\bar{z}_{12}^{\bar{h}_{2}-\bar{h}_{p}-\bar{h}_{1}}L^{-}[O^{-}_{\Delta_{p}}]\nn \\ 
    &+ F(\Delta_{1},\Delta_{2})z_{12}^{h_{p}+h_{1}-h_{2}-1}\bar{z}_{12}^{\bar{h}_{2}-\bar{h}_{p}-\bar{h}_{1}+1}\partial_{\bar{z}_{2}}L^{-}[O^{-}_{\Delta_{p}}]. \label{GeneralGluonOPE}
\end{align}
Consequently, our formula \eqref{generalScaling} tells us that,
\begin{align}
  \Delta_{p}=\Delta_{1}+\Delta_{2}-1 \label{generalGluonScaling}.
\end{align}
Using this in equations \eqref{generalRecursion1.1} and \eqref{generalRecursion1.3} we get,
\begin{align}
    &E(\Delta_{1}+1,\Delta_{2})+E(\Delta_{1},\Delta_{2}+1)\frac{1-\Delta_{1}}{\Delta_{2}}=0 \label{gluonRecursion1}\\
    &E(\Delta_{1}+1,\Delta_{2})+E(\Delta_{1},\Delta_{2}+1)=E(\Delta_{1},\Delta_{2}) \label{gluonRecursion2}
\end{align}
which can be combined to write,
\begin{align}
    &E(\Delta_{1},\Delta_{2}+1)(\Delta_{1}+\Delta_{2}-1)=\Delta_{2}E(\Delta_{1},\Delta_{2})\label{gluonRecursion3}\\
    &E(\Delta_{1}+1,\Delta_{2})(\Delta_{1}+\Delta_{2}-1)=(\Delta_{1}-1)E(\Delta_{1},\Delta_{2}).\label{gluonRecursion4}
\end{align}
Making a change of variable,
\begin{align}
    E(\Delta_{1},\Delta_{2})=(-1)^{\Delta_{1}}\tilde{E}(\Delta_{1},\Delta_{2}) \label{gluonChangeofvariables}
\end{align}
gives us,
\begin{align}
    &\tilde{E}(\Delta_{1},\Delta_{2}+1)(\Delta_{1}+\Delta_{2}-1)=\Delta_{2}\tilde{E}(\Delta_{1},\Delta_{2})\label{gluonRecursion5}\\
    &\tilde{E}(\Delta_{1}+1,\Delta_{2})(1-\Delta_{1}-\Delta_{2})=(\Delta_{1}-1)\tilde{E}(\Delta_{1},\Delta_{2}).\label{gluonRecursion6}
\end{align}
which can be solved using Weiland's theorem and we get,
\begin{align}
    \boxed{E(\Delta_{1},\Delta_{2})=\mathcal{N}_{gluon}(-1)^{\Delta_{1}}B(\Delta_{1}-1,2-\Delta_{1}-\Delta_{2})} \label{gluonOPEcoefficient}
\end{align}
giving us the leading term in our OPE up to a numerical constant,
\begin{align}
    L^{+}[O^{+}_{\Delta_{1}}]L^{-}[O^{-}_{\Delta_{2}}]=-\mathcal{N}_{gluon}B(\Delta_{1}-1,2-\Delta_{1}-\Delta_{2})z_{12}^{\Delta_{1}-1}\bar{z}_{21}^{1-\Delta_{1}}L^{-}[O^{-}_{\Delta_{1}+\Delta_{1}-1}]. \label{IntrinsicGluonOPE}
\end{align}
This is the same result we obtained in \cite{Banerjee:2024hvb} by performing an intrinsic analysis and also by taking the collinear limit of the light-transformed three-point function but by considering only the leading term in the OPE ansatz \eqref{GeneralGluonOPE}. The fact that the scaling dimension or the OPE coefficient did not change in \eqref{IntrinsicGluonOPE} in spite of considering sub-leading terms proves the robustness of the methods used to calculate \eqref{generalScaling} and consequently the above OPE.
\subsubsection{Gravitons and Gluons}
In our general ansatz, we choose $\phi_{\mathfrak{h}_{1}}=G^{+}_{\Delta_{1}}$, $\phi_{\mathfrak{h}_{2}}=O^{-}_{\Delta_{2}}$ and $\phi_{\mathfrak{h}_{p}}=O^{-}_{\Delta_{p}}$. The proposed OPE takes the form,
\begin{align}
     L^{+}[G^{+}_{\Delta_{1}}]L^{-}[O^{-}_{\Delta_{2}}]&=H(\Delta_{1},\Delta_{2})z_{12}^{h_{p}+h_{1}-h_{2}-1}\bar{z}_{12}^{\bar{h}_{2}-\bar{h}_{p}-\bar{h}_{1}}L^{-}[O^{-}_{\Delta_{p}}]\nn \\ 
    &+ J(\Delta_{1},\Delta_{2})z_{12}^{h_{p}+h_{1}-h_{2}-1}\bar{z}_{12}^{\bar{h}_{2}-\bar{h}_{p}-\bar{h}_{1}+1}\partial_{\bar{z}_{2}}L^{-}[O^{-}_{\Delta_{p}}]. \label{GeneralEYMOPE}
\end{align}
OPEs of this form appear when one is considering Einstein-Yang-Mills theory in the bulk \cite{Pate:2019lpp}. Likewise, one can arrive at such OPEs by considering vertices from AdS Witten diagrams involving three-point gluon-graviton vertices as considered in \cite{Bagchi:2023cen}. Now, the formula \eqref{generalScaling} tells us that
\begin{align}
    \Delta_{p}=\Delta_{1}+\Delta_{2} \label{generalEYMscaling}
\end{align}
which corresponds to the correct bulk vertex dimension in EYM theory \cite{Pate:2019lpp}. Following the analysis of the previous sections, one can show that, using \eqref{generalRecursion1.1} and \eqref{generalRecursion1.3}, the OPE coefficient of the leading term satisfies the following recursion relations,
\begin{align}
    &H(\Delta_{1},\Delta_{2}+1)(\Delta_{1}+\Delta_{2}-1)=\Delta_{2}H(\Delta_{1},\Delta_{2})\label{EYMRecursion3}\\
    &H(\Delta_{1}+1,\Delta_{2})(\Delta_{1}+\Delta_{2}-1)=(\Delta_{1}-1)H(\Delta_{1},\Delta_{2}).\label{EYMRecursion4}
\end{align}
The solution to these equations is,
\begin{align}
   \boxed{ H(\Delta_{1},\Delta_{2})=\mathcal{N}_{EYM}(-1)^{\Delta_{1}}B(\Delta_{1}-1,2-\Delta_{1}-\Delta_{2}) }\label{EYMOPECoefficient}
\end{align}
which fixes the leading term of the OPE \eqref{GeneralEYMOPE} as,
\begin{align}
     L^{+}[G^{+}_{\Delta_{1}}]L^{-}[O^{-}_{\Delta_{2}}]=-\mathcal{N}_{EYM}B(\Delta_{1}-1,2-\Delta_{1}-\Delta_{2})z_{12}^{\Delta_{1}}\bar{z}_{21}^{1-\Delta_{1}}L^{-}[O^{-}_{\Delta_{p}}]\nn \\ 
    \label{IntrinsicEYMOPE}
\end{align}
Comparing this with \eqref{GeneralGravitonOPE}, we notice that both OPEs have identical singularity structure, which is well known from \cite{Pate:2019lpp}.
\subsubsection{Marginal gluons}
In \cite{Narayanan:2024qgb}, it was shown that light transforms play an important role when one tries to understand the geometric structure of the space of CFTs dual to asymptotically flat spacetimes. In particular, it helps us construct marginal operators starting with gluon primaries. The analysis was done for ordinary Mellin operators; here we formally extend it for the case of modified Mellin operators. In our general ansatz for OPE, \eqref{GeneralOPEAnsatz}, let us choose $\phi_{\mathfrak{h_{1}}}=O^{-}_{\Delta_{1}}$, $\phi_{\mathfrak{h_{2}}}=O^{+}_{\Delta_{2}}$ and $\phi_{\mathfrak{h_{p}}}=O^{-}_{\Delta_{p}}$ and we obtain,
    \begin{align}
     L^{+}[O^{-}_{\Delta_{1}}]L^{-}[O^{+}_{\Delta_{2}}]&=M(\Delta_{1},\Delta_{2})z_{12}^{h_{p}+h_{1}-h_{2}-1}\bar{z}_{12}^{\bar{h}_{2}-\bar{h}_{p}-\bar{h}_{1}}L^{-}[O^{-}_{\Delta_{p}}]\nn \\ 
    &+ Q(\Delta_{1},\Delta_{2})z_{12}^{h_{p}+h_{1}-h_{2}-1}\bar{z}_{12}^{\bar{h}_{2}-\bar{h}_{p}-\bar{h}_{1}+1}\partial_{\bar{z}_{2}}L^{-}[\phi_{\mathfrak{h}_{p}}]. \label{GeneralMarginalOPE}
\end{align}
It must be understood that in the above equation, none of the operators are marginal unless we set $\Delta=1$. It was shown in \cite{Narayanan:2024qgb} that the only way to construct marginal operators out of light transformed ones is to consider operators of the form $L^{\pm}[O^{\mp}_{\Delta}]$ and then dial $\Delta$ to $1$. Under such a consideration, it is clear that the operator appearing on the right-hand side can never be made marginal, and this is consistent with the findings of \cite{Narayanan:2024qgb}, where it was shown that there are only a finite number of OPEs that will receive contributions from a marginal operator. However, since we are confining ourselves to the space of light-transformed operators, we want to understand whether OPEs of marginal operators can have contributions from non-marginal light-transformed operators or not. Formula \eqref{generalScaling} tells us,
\begin{align}
    \Delta_{p}=\Delta_{1}+\Delta_{2}-1
\end{align}
which further gives us, performing a similar analysis as in the previous cases, 
\begin{align}
    &M(\Delta_{1},\Delta_{2}+1)(\Delta_{1}+\Delta_{2}+1)=(2-\Delta_{2})M(\Delta_{1},\Delta_{2})\label{MarginalRecursion1}\\
    &M(\Delta_{1}+1,\Delta_{2})(1-\Delta_{1}-\Delta_{2})=-(\Delta_{1}+1)M(\Delta_{1},\Delta_{2}). \label{MarginalRecursio2}
\end{align}
These equations are written in a suggestive form so as to be able to compare them to the ones obtained for ordinary gluon \eqref{gluonRecursion3} and \eqref{gluonRecursion4}. The solution to these equations is given by,
\begin{align}
M(\Delta_{1},\Delta_{2})=(-1)^{\Delta_{1}+\Delta_{2}}\mathcal{N}_{Marginal}B(\Delta_{1}+1,2-\Delta_{1}-\Delta_{2}). \label{MarginalCoefficeint}
\end{align}
The leading term in the OPE is,
\begin{align}
     L^{+}[O^{-}_{\Delta_{1}}]L^{-}[O^{+}_{\Delta_{2}}]=\mathcal{N}_{Marginal}B(\Delta_{1}+1,2-\Delta_{1}-\Delta_{2})z_{21}^{\Delta_{1}-3}\bar{z}_{12}^{-1-\Delta_{1}}L^{-}[O^{-}_{\Delta_{p}}].
\end{align}
However, to have marginal operators on the left-hand side, we must take the $\Delta_{1}\rightarrow 1$ and $\Delta_{2}\rightarrow 1$ limit. Now, this corresponds to a consecutive soft limit in the bulk, and since the OPE coefficient is not symmetric in $\Delta_{1}$ and $\Delta_{2}$, we find \footnote{The following form of the OPE with an explicit pole structure in the $\Delta \rightarrow 1$ limit is consistent with \cite{Pate:2019lpp}, although in the context of gluons.} that
\begin{align}
    &\lim_{\Delta_{2}\rightarrow1}\bigg( \lim_{\Delta_{1}\rightarrow1}\bigg)L^{+}[O^{-}_{\Delta_{1}}]L^{-}[O^{+}_{\Delta_{2}}]= \frac{\mathcal{N}_{Marginal}}{1-\Delta_{2}}\frac{1}{z_{21}^{2}\bar{z}_{21}^{2}}L^{-}[O^{-}_{\Delta_{p}}]\\
    &\lim_{\Delta_{1}\rightarrow1}\bigg( \lim_{\Delta_{2}\rightarrow1}\bigg)L^{+}[O^{-}_{\Delta_{1}}]L^{-}[O^{+}_{\Delta_{2}}]= \frac{\mathcal{N}_{Marginal}}{1-\Delta_{1}}\frac{1}{z_{21}^{2}\bar{z}_{21}^{2}}L^{-}[O^{-}_{\Delta_{p}}]
\end{align}
which motivates us to take the symmetrised consecutive soft limit \cite{Klose:2015xoa}. We get,
\begin{align}
    \boxed{\{\lim_{\Delta_{1}\rightarrow1},\lim_{\Delta_{2}\rightarrow1} \}(1-\Delta_{1})(1-\Delta_{2})L^{+}[O^{-}_{\Delta_{1}}]L^{-}[O^{+}_{\Delta_{2}}]=\frac{1}{2}\mathcal{N}_{Marginal}\frac{1-\Delta_{p}}{z_{21}^{2}\bar{z}_{21}^{2}}L^{-}[O^{-}_{\Delta_{p}}]}. \label{MarginalGluonOPE}
\end{align}
This OPE has the characteristic $\frac{1}{|z|^{4}}$ singularity expected from an OPE of marginal operators. One must also note that because we chose to take the symmetrised consecutive soft limit, we obtained a factor of $1-\Delta_{p}$ in the right-hand side, which takes care of the soft limit of the operator appearing on the right-hand side of the OPE. This OPE indicates that there is a contribution from a light-transformed soft operator in the algebra of marginal operators.

\section{Shadow graviton OPE} \label{Shadow graviton OPE}
In this section, motivated by our results with the light-transformed basis, we commence our study of the same analysis for the more familiar shadow basis\cite{Osborn:2012vt, Chang:2022jut, Crawley:2021ivb, Bhattacharyya:2025nfp} given by,
\begin{align}
    S[\phi_{\mathfrak{h}}](u, z,\bar{z})=K_{\mathfrak{h}}\int d^{2}w \frac{\phi_{\mathfrak{h}}(u,w,\bar{w})}{(w-z)^{2-2h}(\bar{w}-\bar{z})^{2-2\bar{h}}} \label{Shadow transform}
\end{align}
where we choose the normalization constant as $K_{\mathfrak{h}}=\frac{\Gamma(2-2\bar{h})}{\pi \Gamma(1-2h)}$.
\subsection{Collinear limit}
 We start by studying the collinear limit of the anti-MHV graviton correlator \eqref{Anti-MHGravitonMellin} in this shadow-transformed basis, 
 \begin{align}
     &\langle S[G^{+}_{\Delta_{1}}] S[G^{+}_{\Delta_{2}}] S[G^{-}_{\Delta_{3}}]\rangle=(-1)^{\Delta_{1}+\Delta_{2}}\mathcal{K}(\Delta_{i})\Gamma(\beta-2)\int d^{2}w_{1} \frac{1}{(w_{1}-z_{1})^{-\Delta_{1}}(\bar{w}_{1}-\bar{z}_{1})^{4-\Delta_{1}}}\nn\\
     &\times \int d^{2}w_{2} \frac{1}{(w_{2}-z_{2})^{-\Delta_{2}}(\bar{w}_{2}-\bar{z}_{2})^{4-\Delta_{2}}}\int d^{2}w_{3} \frac{1}{(w_{3}-z_{3})^{4-\Delta_{3}}(\bar{w}_{3}-\bar{z}_{3})^{-\Delta_{3}}}\delta(w_{12})\delta(w_{23})\bar{w}_{12}^{\Delta_{3}+2}\nn \\
     & \times \bar{w}_{23}^{\Delta_{1}-2}\bar{w}_{31}^{\Delta_{2}-2}\frac{1}{[i(\bar{w}_{1}u_{32}+\bar{w}_{2}u_{13}+\bar{w}_{3}u_{21})]^{\beta-2}} \label{ShadowGravitonAnti-MHV}.
 \end{align}
where $\mathcal{K}(\Delta_{i})=K_{\mathfrak{h}_{1}}K_{\mathfrak{h}_{2}}K_{\mathfrak{h}_{3}}=\frac{\Gamma(4-\Delta_{1})\Gamma(4-\Delta_{2})\Gamma(-\Delta_{3})}{\pi^{3}\Gamma(\Delta_{1}+1)\Gamma(\Delta_{2}+1)\Gamma(\Delta_{3}-3)}$. We perform the integration over $w_{2}$ and $w_{3}$ and expand the denominator in the last line about $u_{23}=0$. We then have,
\begin{align}
      &\langle S[G^{+}_{\Delta_{1}}] S[G^{+}_{\Delta_{2}}] S[G^{-}_{\Delta_{3}}]\rangle=(-1)^{\Delta_{1}+\Delta_{2}}\mathcal{K}(\Delta_{i})\frac{\Gamma(\beta-2)}{(iu_{13})^{\beta-2}} \nn\\
      &\times \int d^{2}w_{1} \frac{1}{(w_{1}-z_{1})^{-\Delta_{1}}(\bar{w}_{1}-\bar{z}_{1})^{4-\Delta_{1}}(w_{1}-z_{3})^{4-\Delta_{3}}(w_{1}-z_{2})^{-\Delta_{2}}}\nn\\ 
      &\times \int d\bar{w}_{2}\frac{\bar{w}_{12}^{\Delta_{3}+2}}{(\bar{w}_{2}-\bar{z}_{2})^{4-\Delta_{2}}} \int d\bar{w}_{3} \ \bar{w}_{23}^{-\Delta_{2}-\Delta_{3}}\bar{w}_{31}^{\Delta_{2}-2} (\bar{w}_{3}-\bar{z}_{3})^{\Delta_{3}}.     
\end{align}
This simplifies to 
\footnote{The integral over $\bar{w}_{3}$ evaluates to,
\begin{align}
     \int d\bar{w}_{3} \ \bar{w}_{23}^{-\Delta_{2}-\Delta_{3}}\bar{w}_{31}^{\Delta_{2}-2} (\bar{w}_{3}-\bar{z}_{3})^{\Delta_{3}}= 
     -(-1)^{\Delta_{3}}B(\Delta_{2}-1,1-\Delta_{2}-\Delta_{3})
     \frac{(\bar{w}_{1}-\bar{z}_{3})^{\Delta_{2}+\Delta_{3}-1}(\bar{w}_{2}-\bar{z}_{3})^{1-\Delta_{2}}}{\bar{w}_{12}^{\Delta_{3}+1}}\nn.
\end{align}
}
\begin{align}
    &\langle S[G^{+}_{\Delta_{1}}] S[G^{+}_{\Delta_{2}}] S[G^{-}_{\Delta_{3}}]\rangle=-(-1)^{\beta}B(\Delta_{2}-1,1-\Delta_{2}-\Delta_{3})\mathcal{K}(\Delta_{i})\frac{\Gamma(\beta-2)}{(iu_{13})^{\beta-2}} \nn\\
      &\times \int d^{2}w_{1} \frac{1}{(w_{1}-z_{1})^{-\Delta_{1}}(\bar{w}_{1}-\bar{z}_{1})^{4-\Delta_{1}}(\bar{w}_{1}-\bar{z}_{3})^{1-\Delta_{2}-\Delta_{3}}(w_{1}-z_{3})^{4-\Delta_{3}}(w_{1}-z_{2})^{-\Delta_{2}}}\nn\\ 
      &\times \int d\bar{w}_{2}(\bar{w}_{2}-\bar{z}_{2})^{4-\Delta_{2}}(\bar{w}_{2}-\bar{z}_{3})^{1-\Delta_{2}}\bar{w}_{12}. \label{ShadowGravitonCollinear1}
\end{align}
 Performing the integration over $\bar{w}_{2}$ give us, 
 \begin{align}
     \int d\bar{w}_{2}(\bar{w}_{2}-\bar{z}_{2})^{4-\Delta_{2}}(\bar{w}_{2}-\bar{z}_{3})^{1-\Delta_{2}}\bar{w}_{12}= \frac{(1-\bar{z}_{2}^{-\Delta_{2}}(-\bar{z}_{3})^{-\Delta_{2}}(-\bar{z}_{2})^{\Delta_{2}}\bar{z}_{3}^{\Delta_{2}})}{\bar{z}_{23}^{2}(\Delta_{2}-2)(\Delta_{2}-3)}(\bar{w}_{1}-\bar{z}_{3}+(\Delta_{2}-2)\bar{z}_{32}).\label{subleadingZbar}
 \end{align}
 Now we note that \cite{Apostol:105425, Ahlfors1966}, 
 \begin{equation}
     (1-\bar{z}_{2}^{-\Delta_{2}}(-\bar{z}_{3})^{-\Delta_{2}}(-\bar{z}_{2})^{\Delta_{2}}\bar{z}_{3}^{\Delta_{2}})=1-e^{2\pi i \Delta_{2} (n(e^{i\pi}, z_{2})-n(e^{i\pi}, z_{3}))} \label{branchCut}
 \end{equation}
 where
 \begin{equation}
n(e^{i\pi},z_{i})=
    \begin{cases}
       0 \quad \textbf{if} \quad -\pi < \pi + \mathrm{arg}(z_{i}) \le \pi\\
       1 \quad \textbf{if} \quad -2\pi < \pi + \mathrm{arg}(z_{i}) \le -\pi \\
       -1 \quad \textbf{if} \quad \pi < \pi + \mathrm{arg}(z_{i}) \le 2\pi.
    \end{cases}
\end{equation}
 So, for the sake of brevity, let us define,
 \begin{align}
     P(\Delta_{2},\bar{z}_{2},\bar{z}_{3})=1-e^{2\pi i \Delta_{2} (n(e^{i\pi}, z_{2})-n(e^{i\pi}, z_{3}))}
 \end{align}
and note that \eqref{branchCut} implies $P(\Delta_{2},\bar{z}_{2},\bar{z}_{3})$ is a numerical factor yet we chose to label it with $\bar{z}_{2}$ and $\bar{z}_{3}$ because its value depends on how these points approach each other, with reference to branch cuts.
 Lastly, let us note that in the $w_{1}$ integral,
\begin{align}
    (w_{1}-z_{3})^{\Delta_{3}-4}(w_{1}-z_{2})^{\Delta_{2}}\approx (w_{1}-z_{3})^{\Delta_{2}+\Delta_{3}-4}-\Delta_{2}z_{23}(w_{1}-z_{3})^{\Delta_{2}+\Delta_{3}-5}.\label{subleadingZ}
\end{align}
 It is worth mentioning that the rationale behind keeping sub-leading terms in both $\bar{z}_{23}$, \eqref{subleadingZbar}, and $z_{23}$ is that the translation symmetry, in contrast to the light basis, acts on the shadow basis with both $z$ and $\bar{z}$ derivatives, as we show in section \ref{GlobalBMSShadow} which means that while checking invariance of this OPE under translations one would require contributions from both kind of descendants. Finally, we have,
 \begin{align}
      &\langle S[G^{+}_{\Delta_{1}}] S[G^{+}_{\Delta_{2}}] S[G^{-}_{\Delta_{3}}]\rangle=-(-1)^{\beta}B(\Delta_{2}-1,1-\Delta_{2}-\Delta_{3})\mathcal{K}(\Delta_{i})\frac{\Gamma(\beta-2)}{(iu_{13})^{\beta-2}} P(\Delta_{2},\bar{z}_{2},\bar{z}_{3}) \nn\\
      &\times \int d^{2}w_{1} \frac{1}{(w_{1}-z_{1})^{-\Delta_{1}}(\bar{w}_{1}-\bar{z}_{1})^{4-\Delta_{1}}(\bar{w}_{1}-\bar{z}_{3})^{1-\Delta_{2}-\Delta_{3}}}\nn \\
      &  \bigg(\frac{\bar{w}_{1}-\bar{z}_{3}}{(\Delta_{2}-2)(\Delta_{2}-3)\bar{z}_{23}^{2}}+\frac{1}{\bar{z}_{32}(\Delta_{2}-3)}\bigg)\bigg( (w_{1}-z_{3})^{\Delta_{2}+\Delta_{3}-4}-\Delta_{2}z_{23}(w_{1}-z_{3})^{\Delta_{2}+\Delta_{3}-5}\bigg)
 \end{align}
 This gives us four terms,
 \begin{align}
      &\langle S[G^{+}_{\Delta_{1}}] S[G^{+}_{\Delta_{2}}] S[G^{-}_{\Delta_{3}}]\rangle\approx -(-1)^{\beta}B(\Delta_{2}-1,1-\Delta_{2}-\Delta_{3})\mathcal{K}(\Delta_{i})\frac{\Gamma(\beta-2)}{(iu_{13})^{\beta-2}}P(\Delta_{2},\bar{z}_{2},\bar{z}_{3})\nn \\
      &\times\Bigg[\frac{1}{(\Delta_{2}-2)(\Delta_{2}-3)\bar{z}_{23}^{2}}\int d^{2}w_{1} \frac{1}{(w_{1}-z_{1})^{-\Delta_{1}}(\bar{w}_{1}-\bar{z}_{1})^{4-\Delta_{1}}(\bar{w}_{1}-\bar{z}_{3})^{-\Delta_{2}-\Delta_{3}}(w_{1}-z_{3})^{4-\Delta_{2}-\Delta_{3}}}\nn \\
      & - \frac{\Delta_{2}z_{23}}{(\Delta_{2}-2)(\Delta_{2}-3)\bar{z}_{23}^{2}}\int d^{2}w_{1} \frac{1}{(w_{1}-z_{1})^{-\Delta_{1}}(\bar{w}_{1}-\bar{z}_{1})^{4-\Delta_{1}}(\bar{w}_{1}-\bar{z}_{3})^{-\Delta_{2}-\Delta_{3}}(w_{1}-z_{3})^{5-\Delta_{2}-\Delta_{3}}}\nn \\
      & +\frac{1}{(\Delta_{2}-3)\bar{z}_{32}}\int d^{2}w_{1} \frac{1}{(w_{1}-z_{1})^{-\Delta_{1}}(\bar{w}_{1}-\bar{z}_{1})^{4-\Delta_{1}}(\bar{w}_{1}-\bar{z}_{3})^{1-\Delta_{2}-\Delta_{3}}(w_{1}-z_{3})^{4-\Delta_{2}-\Delta_{3}}}\nn \\
      &-\frac{\Delta_{2}z_{23}}{(\Delta_{2}-3)\bar{z}_{32}}\int d^{2}w_{1} \frac{1}{(w_{1}-z_{1})^{-\Delta_{1}}(\bar{w}_{1}-\bar{z}_{1})^{4-\Delta_{1}}(\bar{w}_{1}-\bar{z}_{3})^{1-\Delta_{2}-\Delta_{3}}(w_{1}-z_{3})^{5-\Delta_{2}-\Delta_{3}}}\Bigg]. \label{ShadowCollinearOPEIntermediate}
 \end{align}
Now, to obtain the OPE we must compare the right-hand side of the above expression with the two-point function of shadow-transformed gravitons, 
\begin{align}
    &\langle S[G^{+}_{\Delta_{1}}](z_1, \bar{z}_1, u_1)S[G^{-}_{\Delta_{p}}(z_3, \bar{z}_3, u_3)]\rangle =\frac{\Gamma(\Delta_{1}+\Delta_{p}-2)}{\pi^{2}(iu_{31})^{\Delta_{1}+\Delta_{p}-2}}\frac{\Gamma(4-\Delta_{1})\Gamma(-\Delta_{p})}{\Gamma(\Delta_{1}+1)(\Delta_{p}-3)}\nn \\
    & \times \int d^{2}w_{1} \frac{1}{(w_{1}-z_{1})^{-\Delta_{1}}(\bar{w}_{1}-\bar{z}_{1})^{4-\Delta_{1}}(\bar{w}_{1}-\bar{z}_{3})^{-\Delta_{p}}(w_{1}-z_{3})^{4-\Delta_{p}}} \label{ShadowGraviton2pt}.
\end{align}
 Comparing this expression of the two-point function with the right-hand side of \eqref{ShadowCollinearOPEIntermediate} we can write,
 \begin{align}
      &\langle S[G^{+}_{\Delta_{1}}] S[G^{+}_{\Delta_{2}}] S[G^{-}_{\Delta_{3}}]\rangle\approx -\frac{\Gamma(4-\Delta_{2})\Gamma(-\Delta_{3})\Gamma(\Delta_{p}-3)}{\pi\Gamma(\Delta_{2}+1)\Gamma(\Delta_{3}-3)\Gamma(-\Delta_{p})}B(\Delta_{2}-1,1-\Delta_{p})P(\Delta_{2},\bar{z}_{2},\bar{z}_{3}) \nn \\
      &\times \bigg[\frac{1}{(\Delta_{2}-2)(\Delta_{2}-3)\bar{z}_{23}^{2}}\langle S[G^{+}_{\Delta_{1}}]S[G^{-}_{\Delta_{p}}]\rangle +\frac{\Delta_{2}z_{23}}{(\Delta_{p}-4)(\Delta_{2}-2)(\Delta_{2}-3)\bar{z}_{23}^{2}}\partial_{z_{3}}\langle S[G^{+}_{\Delta_{1}}]S[G^{-}_{\Delta_{p}}]\rangle \nn \\
      &  -\frac{1}{(\Delta_{2}-3)\Delta_{p}\bar{z}_{32}} \partial_{\bar{z}_{3}}\langle S[G^{+}_{\Delta_{1}}]S[G^{-}_{\Delta_{p}}]\rangle  -\frac{\Delta_{2}z_{23}}{(\Delta_{2}-3)\Delta_{p}(\Delta_{p}+1)\bar{z}_{32}}\partial_{z_{3}} \partial_{\bar{z}_{3}}\langle S[G^{+}_{\Delta_{1}}]S[G^{-}_{\Delta_{p}}]\rangle\bigg]
 \end{align}
 where we used the expanded form $\mathcal{K}(\Delta_{i})$ and used $\Delta_{p}=\Delta_{1}+\Delta_{2}$. We can read off the OPE as,
 \begin{align}
     S[G^{+}_{\Delta_{2}}] S[G^{-}_{\Delta_{3}}]&=-\frac{P(\Delta_{2},\bar{z}_{2},\bar{z}_{3})\Gamma(\Delta_{p}-3)\Delta_{p}}{\Gamma(\Delta_{2}+1)\Gamma(\Delta_{3}-3)\mathrm{sin}(\pi \Delta_{2})}\frac{1}{\bar{z}_{23}^{2}}S[G^{-}_{\Delta_{2}+\Delta_{3}}]\nn \\
     &-\frac{P(\Delta_{2},\bar{z}_{2},\bar{z}_{3})\Gamma(\Delta_{p}-4)\Delta_{p}}{\Gamma(\Delta_{2})\Gamma(\Delta_{3}-3)\mathrm{sin}(\pi \Delta_{2})}\frac{z_{23}}{\bar{z}_{23}^{2}}\partial_{z_{3}}S[G^{-}_{\Delta_{2}+\Delta_{3}}] \nn\\
     &-\frac{P(\Delta_{2},\bar{z}_{2},\bar{z}_{3})\Gamma(3-\Delta_{2})\Gamma(\Delta_{p}-3)\Gamma(\Delta_{2}-1)}{\pi\Gamma(\Delta_{2}+1)\Gamma(\Delta_{3}-3)}\frac{1}{\bar{z}_{23}}\partial_{\bar{z}_{3}}S[G^{+}_{\Delta_{p}}]\nn \\
     &-\frac{P(\Delta_{2},\bar{z}_{2},\bar{z}_{3})\Gamma(3-\Delta_{2})\Gamma(\Delta_{p}-4)\Gamma(\Delta_{2}-1)}{\pi\Gamma(\Delta_{2})\Gamma(\Delta_{3}-3)}\frac{z_{23}}{\bar{z}_{23}}\partial_{z_{3}}\partial_{\bar{z}_{3}}S[G^{+}_{\Delta_{p}}] \label{ShadowGravitonOPECollinear}
 \end{align}
The shadow transform, unlike the light transform, affects both the coordinates $z$ and $\bar{z}$, which is why it is natural that sub-leading terms in the OPE of shadow operators have both holomorphic and anti-holomorphic contributions, and that is why we have three sub-leading terms instead of one. It is worth noting that this OPE is not invariant under $z\rightleftharpoons \bar{z}$; however, that can be resolved by deriving the OPE of $S[G^{-}_{\Delta_{2}}] S[G^{+}_{\Delta_{3}}]$, from the shadow transformed MHV three-point function $\langle S[G^{-}_{\Delta_{1}}]S[G^{-}_{\Delta_{2}}]S[G^{+}_{\Delta_{3}}]\rangle$, and then interchanging $\Delta_{2}$ and $\Delta_{3}$ followed by adding this contribution to \eqref{ShadowGravitonOPECollinear}\footnote{We thank Monica Pate for pointing this out.}. As before, we want to ensure that the OPE that we derived using the collinear limit satisfies the translation symmetries. To that end, we need to understand how translations act on the shadow operators. We explicitly derive them in the next section.

\section{Symmetries of the shadow-graviton OPE} \label{GlobalBMSShadow}

In this section, we study how the shadow transformed Carroll primaries transform under the Poincar\'e transformations. Utilising these, we will finally check the symmetry properties of the shadow graviton operators \eqref{ShadowGravitonOPECollinear}. For the present set of analyses, we will treat $z$ and $\bar{z}$ as independent complex coordinates, and after we have our final answer, we can analytically continue back to the \textit{real surface} \cite{DiFrancesco:1997nk}.
\subsection{Transformation properties under global BMS}
\subsubsection*{Translation}
We start by looking at how $\delta_{M_{00}}$ acts on the shadow operator. We closely follow the analysis presented in \cite{Banerjee:2024hvb}. Starting with 
\begin{equation}
    \delta_{M_{00}}[\phi_{\mathfrak{h}}]= K_{\mathfrak{h}}\int d^{2}w \frac{\delta_{M_{00}}\phi_{\mathfrak{h}}}{(w-z)^{2-2h}(\Bar{w}-\Bar{z})^{2-2\Bar{h}}}
\end{equation}
where 
$$
K_{\mathfrak{h}}=\frac{\Gamma(2-2\Bar{h})}{\pi\Gamma(2h-1)}.
$$
Using the fact that $\delta_{M_{00}}\phi_{\mathfrak{h}}=-i\epsilon \phi_{\mathfrak{h}+\frac{1}{2}}$, we get,
\begin{equation}
     \delta_{M_{00}}S[\phi_{\mathfrak{h}}]= -i\epsilon \frac{K_{\mathfrak{h}}}{K_{\mathfrak{h}+\frac{1}{2}}}\frac{\partial_{z}\partial_{\Bar{z}}}{(1-2h)(1-2\Bar{h})}S[\phi_{\mathfrak{h}+\frac{1}{2}}] 
\end{equation}
The factor sitting in front of the derivatives simplifies to $ \frac{K_{\mathfrak{h}}}{K_{\mathfrak{h}+\frac{1}{2}}}=(1-2\Bar{h})(2h-1)$. Thus, we have,
\begin{equation}
    \delta_{M_{00}}S[\phi_{\mathfrak{h}}]=i\epsilon \partial_{z}\partial_{\Bar{z}}S[\phi_{\mathfrak{h}+\frac{1}{2}}] \label{M00ShadowAction}
\end{equation}
Now, for $\delta_{M_{10}}$ one needs to take care of the fact that there is an extra factor of $z$ sitting in the action of $\delta_{M_{10}}$ on a conformal Carroll operator, nevertheless the analysis for the anti-holomorphic part of \eqref{M00ShadowAction} will remain the same. The converse holds for $\delta_{M_{01}}$. After using the residue theorem, we get the following results for the rest of the translations, 
\begin{align}
    &  \delta_{M_{01}}S[\phi_{\mathfrak{h}}]= i\epsilon \partial_{z}\Bigg(\Bar{z}\partial_{\Bar{z}}+1-2\Bar{h}\Bigg) S[\phi_{\mathfrak{h}+\frac{1}{2}}] \label{M01Action}\\
    &\delta_{M_{10}}S[\phi_{\mathfrak{h}}]= i\epsilon\partial_{\Bar{z}}\Bigg(z\partial_{z}+1-2h\Bigg) S[\phi_{\mathfrak{h}+\frac{1}{2}}]\label{M10Action} \\
    & \delta_{M_{11}}S[\phi_{\mathfrak{h}}]= i\epsilon\Bigg(\Bar{z}\partial_{\Bar{z}}+1-2\Bar{h}\Bigg)\Bigg(z\partial_{z}+1-2h\Bigg)S[\phi_{\mathfrak{h}+\frac{1}{2}}]. \label{M11Action}
\end{align}
Similar results were obtained in \cite{Chang:2022jut} by considering celestial conformal primaries, which is expected since the $4D$ translation generators are trivially projected onto the global part of $3D$ super-translations.
\subsubsection*{Lorentz transformation}
Now, let us look at how global conformal Carroll transformations, which are isomorphic to Lorentz transformations, act on shadow operators. We closely follow the analysis from the previous section to simplify our results. The translations along $z$ and $\Bar{z}$ direction act on the shadow operators in the following way,
\begin{align}
    &\delta_{L_{-1}}S[\phi_{\mathfrak{h}}]= \partial_{z}S[\phi_{\mathfrak{h}}] \label{L-1Action}\\
    &\delta_{\Bar{L}_{-1}}S[\phi_{\mathfrak{h}}]= \partial_{\Bar{z}}S[\phi_{\mathfrak{h}}]. \label{BarL-1Action}
\end{align}
Next, we look at how the generators of scale transformation act on these operators,
\begin{align}
    &\delta_{L_{0}}S[\phi_{\mathfrak{h}}]=\bigg(z\partial_{z}+(1-h)\bigg)S[\phi_{\mathfrak{h}}]+\frac{i\epsilon u}{2}\partial_{z}\partial_{\Bar{z}}S[\phi_{\mathfrak{h}+\frac{1}{2}}] \label{L0Action}\\
    &\delta_{\Bar{L}_{0}}S[\phi_{\mathfrak{h}}]=\bigg(\Bar{z}\partial_{\Bar{z}}+(1-\Bar{h})\bigg)S[\phi_{\mathfrak{h}}] +\frac{i\epsilon u}{2}\partial_{z}\partial_{\Bar{z}}S[\phi_{\mathfrak{h}+\frac{1}{2}}]. \label{LBar0Action}
\end{align}
Lastly, we look at special conformal transformations,
\begin{align}
    &\delta_{L_{1}}S[\phi_{\mathfrak{h}}]=\bigg(z^{2}\partial_{z}+2(1-h)z\bigg)S[\phi_{\mathfrak{h}}]+i\epsilon u \partial_{\Bar{z}}
\Bigg(\partial_{z}+1-2h \Bigg)S[\phi_{\mathfrak{h}+\frac{1}{2}}] \label{L1Action} \\
& \delta_{\Bar{L}_{1}}S[\phi_{\mathfrak{h}}]=\bigg(\Bar{z}^{2}\partial_{\Bar{z}}+2(1-\Bar{h})\Bar{z} \bigg)S[\phi_{\mathfrak{h}}] + i\epsilon u \partial_{z}\Bigg(\partial_{\Bar{z}}+1-2\Bar{h}\Bigg) S[\phi_{\mathfrak{h}+\frac{1}{2}}]
\end{align}
The above equations immediately tell us that $S[\phi_{\mathfrak{h}}]$ is a primary of weight $(1-h,1-\Bar{h})$ under the $SL(2,C)$ group.
\subsection{Symmetry analysis of shadow OPE}
Equipped with equation \eqref{M00ShadowAction}, we are now in a position to analyse whether the OPE, \eqref{ShadowGravitonOPECollinear}, we obtained by taking the collinear limit of the shadow three-point correlator, satisfies translation symmetry or not. We start by demanding the validity of the following: 
\begin{align}
     \delta_{M_{00}}S[G^{+}_{\Delta_{2}}] S[G^{-}_{\Delta_{3}}]+&S[G^{+}_{\Delta_{2}}]\delta_{M_{00}} S[G^{-}_{\Delta_{3}}] \stackrel{?}{=} -\frac{P(\Delta_{2},\bar{z}_{2},\bar{z}_{3})\Gamma(\Delta_{p}-3)\Delta_{p}}{\Gamma(\Delta_{2}+1)\Gamma(\Delta_{3}-3)\mathrm{sin}(\pi \Delta_{2})}\frac{1}{\bar{z}_{23}^{2}}\delta_{M_{00}}S[G^{-}_{\Delta_{2}+\Delta_{3}}]\nn \\
     &-\frac{P(\Delta_{2},\bar{z}_{2},\bar{z}_{3})\Gamma(\Delta_{p}-4)\Delta_{p}}{\Gamma(\Delta_{2})\Gamma(\Delta_{3}-3)\mathrm{sin}(\pi \Delta_{2})}\frac{z_{23}}{\bar{z}_{23}^{2}}\partial_{z_{3}}\delta_{M_{00}}S[G^{-}_{\Delta_{2}+\Delta_{3}}] \nn\\
     &-\frac{P(\Delta_{2},\bar{z}_{2},\bar{z}_{3})\Gamma(3-\Delta_{2})\Gamma(\Delta_{p}-3)\Gamma(\Delta_{2}-1)}{\pi\Gamma(\Delta_{2}+1)\Gamma(\Delta_{3}-3)}\frac{1}{\bar{z}_{23}}\partial_{\bar{z}_{3}}\delta_{M_{00}}S[G^{+}_{\Delta_{p}}]\nn \\
     &-\frac{P(\Delta_{2},\bar{z}_{2},\bar{z}_{3})\Gamma(3-\Delta_{2})\Gamma(\Delta_{p}-4)\Gamma(\Delta_{2}-1)}{\pi\Gamma(\Delta_{2})\Gamma(\Delta_{3}-3)}\frac{z_{23}}{\bar{z}_{23}}\partial_{z_{3}}\partial_{\bar{z}_{3}}\delta_{M_{00}}S[G^{+}_{\Delta_{p}}] . \label{CollinearShadowOPECheck}
\end{align}
Using \eqref{M00ShadowAction} and the OPE \eqref{ShadowGravitonOPECollinear} to simplify the left-hand side, we get, after some tedious algebra\footnote{We show these calculations in Appendix \ref{Intermediate steps for checking translation symmetry of shadow OPE}.},
\begin{align}
   & -i\epsilon\frac{\Gamma(\Delta_{p}-3)\Delta_{p}}{\Gamma(\Delta_{2}+1)\Gamma(\Delta_{3}-3)\mathrm{sin}(\pi\Delta_{2})}\frac{P(\Delta_{2},\bar{z}_{2},\bar{z}_{3})}{\bar{z}_{23}^{2}}\partial_{z_{3}}\partial_{\bar{z}_{3}} S[G^{-}_{\Delta_{p}+1}] \nn \\
   & -i\epsilon\frac{\Gamma(3-\Delta_{2})\Gamma(\Delta_{p}-3)\Gamma(\Delta_{2}-1)}{\pi \Gamma(\Delta_{2}+1)\Gamma(\Delta_{3}-3)}\frac{P(\Delta_{2},\bar{z}_{2},\bar{z}_{3})}{\bar{z}_{23}}\partial_{z_{3}}\partial_{\bar{z}_{3}}^{2}S[G^{-}_{\Delta_{p}+1}]+ \text{other sub-leading terms}. \label{LHSShadowCollienarOPEM00Check}
\end{align}
Comparing this with the right-hand side of \eqref{CollinearShadowOPECheck}, we see that the OPE of gravitons in the shadow basis satisfies the $\delta_{M_{00}}$ symmetry at the leading order and sub-leading order in $\bar{z}_{23}$. Again, like the light-transformed basis, we required contributions from sub-leading terms to ensure that the symmetry holds at the level of the OPE, and one would require sub$^{2}$-leading terms to ensure that the symmetry holds for the terms we labelled as `other sub-leading terms' in \eqref{LHSShadowCollienarOPEM00Check}. 

\section{Discussion and future directions}
The aim of our program with light transforms is to come up with a basis starting from conformal Carroll (or celestial) operators such that the correlation functions have the correct power law behaviour on the celestial sphere, as is expected from a CFT, and to satisfy global conformal symmetry. The next step is calculating the algebra of operators using the OPE limit of the correlation functions, which is found by taking a collinear limit of the correlators. What is new, then? We observed that in the light-transformed basis, one could, owing to the non-trivial action of translations (global part of the supertranslations) on them, fix the scaling dimension and coefficient (up to a numerical constant) of the leading term in the OPE of these operators. In the `light' of this program, let us summarise the key aspects of the present work as follows:
\begin{itemize}
\item We started with a three-point graviton correlator written in the basis of light-transformed conformal Carroll operators and calculated a putative OPE from it by taking the collinear limit. To verify the authenticity of this OPE, we check that it is invariant under translations, and we learn that to ensure that the symmetry holds at the leading order, one needs to keep the sub-leading order term. 

\item Taking insight from the calculations of section \ref{gravitonSymmetry}, we start with an ansatz for the OPE of light-transformed operators and consider sub-leading terms in that OPE. Using purely symmetry-based arguments, we have been able to find a formula that gives us the scaling dimension of the conformal Carroll operator whose light transform contributes to the leading term in the OPE. Along with this, we have also found the corresponding OPE coefficient.  It is worth mentioning that this formalism can, in principle, be used to fix OPEs of the form $L^{-}[\phi_{\mathfrak{h}_{1}}]L^{+}[\phi_{\mathfrak{h}_{2}}]$ as well. This will completely classify the algebra of these operators. Owing to the fact that $\phi_{\mathfrak{h}}$ are Carroll operators, this opens another avenue for studying the Carroll CFT, which are putative duals to asymptotically flat space, from an intrinsic perspective. 

    \item To verify that the formula for scaling dimension we derived, \eqref{generalScaling}, was correct or not, we put it to the test by calculating the scaling dimension for graviton, gluon and gluon-graviton OPEs and saw that the calculated scaling dimension matches with existing literature. For the graviton and gluon case, the OPEs matched exactly with the ones calculated by taking the collinear limit in the light-transformed basis.
    
   \item It is well established that, in Carroll (or in celestial) CFT, the scaling of the operator appearing in the leading term of the OPE is calculated by looking at the bulk dimension of the three-point vertex in momentum space from which the OPE was calculated. This, along with our results, points us towards the central result of this paper: \textit{in the light-transformed basis, one can calculate the scaling dimension of the operator, intrinsically, in the leading term of the OPE; thereby, one can comment on dimension of the bulk vertex in momentum space and consequently on the possible interactions that one can have in a bulk theory}.   
    \item We further calculate the graviton OPE in the shadow basis and show that the leading term is more singular than the one existing in the literature. However, we found that the OPE respects translation symmetry, and, owing to the democratic nature of the shadow transform, one must keep descendants in both $z$ and $\bar{z}$ to verify that the symmetry holds at the leading order. The fact that the shadow operators have a different singular structure as compared to the usual celestial CFT primaries is something worth understanding since both are part of the same unitary principal series.
    We leave this for future perusal.
    \item  The shadow operator basis is complementary to the ordinary Mellin basis in celestial CFT. The role of shadow operators in the representation theory of Carroll CFT is not well understood to the best of the authors' knowledge. Finding the OPE in shadow basis from symmetry, in the same spirit as light-transformed operators, is something which we want to pursue in future works. This will help one classify the spectrum of operators required for a better comprehension of the boundary CFT.
    \item Since other authors and we are putting considerable effort into understanding the local singularity structure in operator algebra in BMS/ Carroll symmetric theories, it is imperative that further studies should be done in understanding the associativity of these operator algebras (look at \cite{Costello:2022upu, Ren:2022sws} for comments on associativity of OPE in celestial CFT). As in traditional $2D$ CFT, this lies inherently in the crossing symmetry of 4-point functions. An analysis for lower-dimensional BMS invariant theories was done in \cite{Bagchi:2016geg, Bagchi:2017cpu}.
\section*{Acknowledgement}
SB thanks Sayali Bhatkar for her useful comments on the work and for past collaborations that have led to these results. SB is grateful to IIT Kanpur for their hospitality during his visit and is thankful to Arjun Bagchi, Diptarka Das and Amartya Saha for insightful discussions while this work was in progress. SB thanks Daniel Grumiller for hosting him at TU Wien and for various interesting discussions while this work was in progress. The authors would like to thank Sruthi Narayanan and Monica Pate for their comments and suggestions. The authors would like to thank all the participants of ``Holography, strings and other fun things II" for engaging discussions and inputs on this work while it was in a premature stage. The grants that support the research of RB are MTR/ 2022/000795 from ANRF, India, the CDRF and OPERA grants from BITS Pilani, and the Indo-Austria bilateral research grant DST/IC/Austria/P-9/2021.
\end{itemize}
\appendix
\section{Translation symmetry constraint on the general structure of OPE} \label{translationSymmetry}
In this section, we explicitly show the calculations that lead to the results given in Section \ref{generalOPEScaling}.
We start by re-writing the action $\delta_{M_{01}}$ on light-transformed Carroll primaries,
\begin{align}
    \delta_{M_{01}}L^{+}[\phi_{\mathfrak{h}}]=-i\epsilon\frac{\bar{z}\partial_{z}}{1-2h}L^{+}[\phi_{\mathfrak{h}+\frac{1}{2}}] \qquad \text{and} \qquad  \delta_{M_{01}}L^{-}[\phi_{\mathfrak{h}}]=-i\epsilon\bigg(\frac{\bar{z}\partial_{z}}{1-2\bar{h}}+1\bigg)L^{-}[\phi_{\mathfrak{h}+\frac{1}{2}}] \label{M01LightAction}.
    \end{align}
After acting on the OPE \eqref{GeneralOPEAnsatz} and simplifying the expression, we get,
\begin{align}
    &i\epsilon \Big[\frac{C(\mathfrak{h}_{1}+\frac{1}{2},\mathfrak{h}_{2})}{1-2h_{1}}(h_{p}+h_{1}-h_{2})\bar{z}_{1}+\frac{C(\mathfrak{h}_{1},\mathfrak{h}_{2}+\frac{1}{2})}{1-2\bar{h}_{2}}(\bar{h}_{2}-\bar{h}_{p}-\bar{h}_{1})\bar{z}_{2}\Big]z_{12}^{h_{p}+h_{1}-h_{2}-1}\bar{z}_{12}^{\bar{h}_{2}-\bar{h}_{p}-\bar{h}_{1}-1}L^{-}[\phi_{\mathfrak{h}_{p}+\frac{1}{2}}] \nn \\
    & +i\epsilon \Big[\frac{D(\mathfrak{h}_{1}+\frac{1}{2},\mathfrak{h}_{2})}{1-2h_{1}}(h_{p}+h_{1}-h_{2})\bar{z}_{1}-\frac{C(\mathfrak{h}_{1},\mathfrak{h}_{2}+\frac{1}{2})}{1-2\bar{h}_{2}}\bar{z}_{2}\nn \\
    &+\frac{D(\mathfrak{h}_{1},\mathfrak{h}_{2}+\frac{1}{2})}{1-2\bar{h}_{2}}(\bar{h}_{2}-\bar{h}_{p}-\bar{h}_{1}+1)\bar{z}_{2} \Big]z_{12}^{h_{p}+h_{1}-h_{2}-1}\bar{z}_{12}^{\bar{h}_{2}-\bar{h}_{p}-\bar{h}_{1}}\partial_{\bar{z}_{2}}L^{-}[\phi_{\mathfrak{h}_{p}+\frac{1}{2}}]\nn \\
    & -i\epsilon \frac{D(\mathfrak{h},\mathfrak{h}_{2}+\frac{1}{2})}{1-2\bar{h}_{2}}z_{12}^{h_{p}+h_{1}-h_{2}-1}\bar{z}_{12}^{\bar{h}_{2}-\bar{h}_{p}-\bar{h}_{1}+1}\bar{z}_{2}\partial^{2}_{\bar{z}_{2}}L^{-}[\phi_{\mathfrak{h}_{p}+\frac{1}{2}}]\nn \\
    &-i\epsilon C(\mathfrak{h},\mathfrak{h}_{2}+\frac{1}{2})z_{12}^{h_{p}+h_{1}-h_{2}-1}\bar{z}_{12}^{\bar{h}_{2}-\bar{h}_{p}-\bar{h}_{1}}L^{-}[\phi_{\mathfrak{h}_{p}+\frac{1}{2}}]-i\epsilon D(\mathfrak{h},\mathfrak{h}_{2}+\frac{1}{2})z_{12}^{h_{p}+h_{1}-h_{2}-1}\bar{z}_{12}^{\bar{h}_{2}-\bar{h}_{p}-\bar{h}_{1}+1}\partial_{\bar{z}_{2}}L^{-}[\phi_{\mathfrak{h}_{p}+\frac{1}{2}}] \nn \\
    &=-i\epsilon C(\mathfrak{h}_{1},\mathfrak{h}_{2})z_{12}^{h_{p}+h_{1}-h_{2}-1}\bar{z}_{12}^{\bar{h}_{2}-\bar{h}_{p}-\bar{h}_{1}}\bigg(\frac{\bar{z}_{2}\partial_{\bar{z}_{2}}}{1-2\bar{h}_{2}} +1\bigg)L^{-}[\phi_{\mathfrak{h}_{p}+\frac{1}{2}}] \nn \\
    & -i\epsilon D(\mathfrak{h}_{1},\mathfrak{h}_{2})z_{12}^{h_{p}+h_{1}-h_{2}-1}\bar{z}_{12}^{\bar{h}_{2}-\bar{h}_{p}-\bar{h}_{1}+1}\partial_{\bar{z}_{2}}\bigg(\frac{\bar{z}_{2}\partial_{\bar{z}_{2}}}{1-2\bar{h}_{p}} +1\bigg)L^{-}[\phi_{\mathfrak{h}_{p}+\frac{1}{2}}].
\end{align}
In the above equation we simplify the expression on the left-hand side by using $\bar{z}_{1}=\bar{z}_{12}+\bar{z}_{2}$ and on the right-hand side we simplify the second term by performing the differentiation and get, 
\begin{align}
    &i\epsilon \Big[\frac{C(\mathfrak{h}_{1}+\frac{1}{2},\mathfrak{h}_{2})}{1-2h_{1}}(h_{p}+h_{1}-h_{2})+\frac{C(\mathfrak{h}_{1},\mathfrak{h}_{2}+\frac{1}{2})}{1-2\bar{h}_{2}}(\bar{h}_{2}-\bar{h}_{p}-\bar{h}_{1})\Big]z_{12}^{h_{p}+h_{1}-h_{2}-1}\bar{z}_{12}^{\bar{h}_{2}-\bar{h}_{p}-\bar{h}_{1}-1}\bar{z}_{2}L^{-}[\phi_{\mathfrak{h}_{p}+\frac{1}{2}}] \nn \\
    &+ i\epsilon \Big[\frac{D(\mathfrak{h}_{1}+\frac{1}{2},\mathfrak{h}_{2})}{1-2h_{1}}(h_{p}+h_{1}-h_{2})-\frac{C(\mathfrak{h}_{1},\mathfrak{h}_{2}+\frac{1}{2})}{1-2\bar{h}_{2}}\nn \\
    &+\frac{D(\mathfrak{h}_{1},\mathfrak{h}_{2}+\frac{1}{2})}{1-2\bar{h}_{2}}(\bar{h}_{2}-\bar{h}_{p}-\bar{h}_{1}+1) \Big]z_{12}^{h_{p}+h_{1}-h_{2}-1}\bar{z}_{12}^{\bar{h}_{2}-\bar{h}_{p}-\bar{h}_{1}}\bar{z}_{2}\partial_{\bar{z}_{2}}L^{-}[\phi_{\mathfrak{h}_{p}+\frac{1}{2}}]\nn \\
    &+i\epsilon \bigg[\frac{C(\mathfrak{h}_{1}+\frac{1}{2},\mathfrak{h}_{2})}{1-2h_{1}}(h_{p}+h_{1}-h_{2})-C(\mathfrak{h}_{1},\mathfrak{h}_{2}+\frac{1}{2})\bigg]z_{12}^{h_{p}+h_{1}-h_{2}-1}\bar{z}_{12}^{\bar{h}_{2}-\bar{h}_{p}-\bar{h}_{1}}L^{-}[\phi_{\mathfrak{h}_{p}+\frac{1}{2}}]\nn \\
    &+i\epsilon \bigg[\frac{D(\mathfrak{h}_{1}+\frac{1}{2},\mathfrak{h}_{2})}{1-2h_{1}}(h_{p}+h_{1}-h_{2})-D(\mathfrak{h}_{1},\mathfrak{h}_{2}+\frac{1}{2})\bigg]z_{12}^{h_{p}+h_{1}-h_{2}-1}\bar{z}_{12}^{\bar{h}_{2}-\bar{h}_{p}-\bar{h}_{1}+1}\partial_{\bar{z}_{2}}L^{-}[\phi_{\mathfrak{h}_{p}+\frac{1}{2}}]\nn \\
    &-i\epsilon \frac{D(\mathfrak{h}_{1},\mathfrak{h}_{2}+\frac{1}{2})}{1-2\bar{h}_{2}}z_{12}^{h_{p}+h_{1}-h_{2}-1}\bar{z}_{12}^{\bar{h}_{2}-\bar{h}_{p}-\bar{h}_{1}+1}\bar{z}_{2}\partial^{2}_{\bar{z}_{2}}L^{-}[\phi_{\mathfrak{h}_{p}+\frac{1}{2}}] \nn \\
    &=-i\epsilon C(\mathfrak{h}_{1},\mathfrak{h}_{2})z_{12}^{h_{p}+h_{1}-h_{2}-1}\bar{z}_{12}^{\bar{h}_{2}-\bar{h}_{p}-\bar{h}_{1}}\bigg(\frac{\bar{z}_{2}\partial_{\bar{z}_{2}}}{1-2\bar{h}_{2}} +1\bigg)L^{-}[\phi_{\mathfrak{h}_{p}+\frac{1}{2}}] \nn \\
    & -i\epsilon D(\mathfrak{h}_{1},\mathfrak{h}_{2})z_{12}^{h_{p}+h_{1}-h_{2}-1}\bar{z}_{12}^{\bar{h}_{2}-\bar{h}_{p}-\bar{h}_{1}+1}\bigg(\frac{2-2\bar{h}_{p}}{1-2\bar{h}_{p}}\partial_{\bar{z}_{2}}+\bar{z}_{2}\partial^{2}_{\bar{z}_{2}}\bigg)L^{-}[\phi_{\mathfrak{h}_{p}+\frac{1}{2}}] .\label{M01generalOPE}
\end{align}
Comparing similar powers of $z_{12}$ and $\bar{z}_{12}$ in the leading order gives us \eqref{generalRecursion3} along with \eqref{generalRecursion1} and \eqref{generalRecursion2}. However, let us observe at this stage that the third term in rectangular braces will never receive a correction, even if we were to consider sub-leading terms, this observation directly implies \eqref{generalRecursion4}. 

Next we use the $\delta_{M_{10}}$ symmetry, its action on the light-transformed fields are, 
\begin{align}
    \delta_{M_{10}}L^{+}[\phi_{\mathfrak{h}}]=-i\epsilon \bigg( \frac{z\partial_{z}}{1-2h}+1\bigg)L^{+}[\phi_{\mathfrak{h}+\frac{1}{2}}] \qquad \text{and} \qquad  \delta_{M_{10}}L^{-}[\phi_{\mathfrak{h}}]=-i\epsilon  \frac{z\partial_{\bar{z}}}{1-2h}L^{-}[\phi_{\mathfrak{h}+\frac{1}{2}}] \label{M10LightAction}.
\end{align}
Using this on both the sides of the OPE \eqref{GeneralOPEAnsatz} we get, after some simplification, 
\begin{align}
    &i\epsilon \Big[\frac{C(\mathfrak{h}+\frac{1}{2},\mathfrak{h}_{2})}{1-2h_{1}}(h_{p}+h_{1}-h_{2})z_{1}+\frac{C(\mathfrak{h}_{1},\mathfrak{h}_{2}+\frac{1}{2})}{1-2\bar{h}_{2}}(\bar{h}_{2}-\bar{h}_{p}-\bar{h}_{1})z_{2} \Big ]z_{12}^{h_{p}+h_{1}-h_{2}-1}\bar{z}_{12}^{\bar{h}_{2}-\bar{h}_{p}-\bar{h}_{1}-1}L^{-}[\phi_{\mathfrak{h}_{p}+\frac{1}{2}}] \nn \\
   & +i\epsilon  \Big[ \frac{D(\mathfrak{h}+\frac{1}{2},\mathfrak{h}_{2})}{1-2h_{1}}(h_{p}+h_{1}-h_{2})z_{1} - \frac{C(\mathfrak{h}_{1},\mathfrak{h}_{2}+\frac{1}{2})}{1-2\bar{h}_{2}}z_{2}\nn \\
   &+\frac{D(\mathfrak{h}_{1},\mathfrak{h}_{2}+\frac{1}{2})}{1-2\bar{h}_{2}}(\bar{h}_{2}-\bar{h}_{p}-\bar{h}_{1}+1)z_{2}\Big]z_{12}^{h_{p}+h_{1}-h_{2}-1}\bar{z}_{12}^{\bar{h}_{2}-\bar{h}_{p}-\bar{h}_{1}}\partial_{\bar{z}_{2}}L^{-}[\phi_{\mathfrak{h}_{p}+\frac{1}{2}}] \nn \\
   &+i\epsilon C(\mathfrak{h}_{1}+\frac{1}{2},\mathfrak{h}_{2})z_{12}^{h_{p}+h_{1}-h_{2}}\bar{z}_{12}^{\bar{h}_{2}-\bar{h}_{p}-\bar{h}_{1}-1}L^{-}[\phi_{\mathfrak{h}_{p}+\frac{1}{2}}] \nn\\
   &+i\epsilon D(\mathfrak{h}_{1}+\frac{1}{2},\mathfrak{h}_{2})z_{12}^{h_{p}+h_{1}-h_{2}}\bar{z}_{12}^{\bar{h}_{2}-\bar{h}_{p}-\bar{h}_{1}}\partial_{\bar{z}_{2}}L^{-}[\phi_{\mathfrak{h}_{p}+\frac{1}{2}}] \nn\\
   &-i\epsilon \frac{D(\mathfrak{h}_{1},\mathfrak{h}_{2}+\frac{1}{2})}{1-2\bar{h}_{2}}z_{12}^{h_{p}+h_{1}-h_{2}-1}\bar{z}_{12}^{\bar{h}_{2}-\bar{h}_{p}-\bar{h}_{1}+1}z_{2}\partial^{2}_{\bar{z}_{2}}L^{-}[\phi_{\mathfrak{h}_{p}+\frac{1}{2}}]\nn \\
   &=-i\epsilon C(\mathfrak{h}_{1},\mathfrak{h}_{2})z_{12}^{h_{p}+h_{1}-h_{2}-1}\bar{z}_{12}^{\bar{h}_{2}-\bar{h}_{p}-\bar{h}_{1}}\bigg( \frac{z_{2}\partial_{\bar{z}_{2}}}{1-2\bar{h}_{p}}\bigg)L^{-}[\phi_{\mathfrak{h}_{p}+\frac{1}{2}}]\nn \\
   &-i\epsilon D(\mathfrak{h}_{1},\mathfrak{h}_{2})z_{12}^{h_{p}+h_{1}-h_{2}-1}\bar{z}_{12}^{\bar{h}_{2}-\bar{h}_{p}-\bar{h}_{1}+1}\bigg( \frac{z_{2}\partial_{\bar{z}_{2}}}{1-2\bar{h}_{p}}\bigg)\partial_{\bar{z}_{2}}L^{-}[\phi_{\mathfrak{h}_{p}+\frac{1}{2}}].
\end{align}
Again, re-writing $z_{1}=z_{12}+z_{2}$, we get,
\begin{align}
    &i\epsilon \Big[\frac{C(\mathfrak{h}+\frac{1}{2},\mathfrak{h}_{2})}{1-2h_{1}}(h_{p}+h_{1}-h_{2})+\frac{C(\mathfrak{h}_{1},\mathfrak{h}_{2}+\frac{1}{2})}{1-2\bar{h}_{2}}(\bar{h}_{2}-\bar{h}_{p}-\bar{h}_{1}) \Big ]z_{12}^{h_{p}+h_{1}-h_{2}-1}\bar{z}_{12}^{\bar{h}_{2}-\bar{h}_{p}-\bar{h}_{1}-1}z_{2}L^{-}[\phi_{\mathfrak{h}_{p}+\frac{1}{2}}] \nn \\
   & +i\epsilon  \Big[ \frac{D(\mathfrak{h}+\frac{1}{2},\mathfrak{h}_{2})}{1-2h_{1}}(h_{p}+h_{1}-h_{2}) - \frac{C(\mathfrak{h}_{1},\mathfrak{h}_{2}+\frac{1}{2})}{1-2\bar{h}_{2}}\nn \\
   &+\frac{D(\mathfrak{h}_{1},\mathfrak{h}_{2}+\frac{1}{2})}{1-2\bar{h}_{2}}(\bar{h}_{2}-\bar{h}_{p}-\bar{h}_{1}+1)\Big]z_{12}^{h_{p}+h_{1}-h_{2}-1}\bar{z}_{12}^{\bar{h}_{2}-\bar{h}_{p}-\bar{h}_{1}}z_{2}\partial_{\bar{z}_{2}}L^{-}[\phi_{\mathfrak{h}_{p}+\frac{1}{2}}] \nn \\
  & +i\epsilon \Big[\frac{C(\mathfrak{h}+\frac{1}{2},\mathfrak{h}_{2})}{1-2h_{1}}(h_{p}+h_{1}-h_{2})+C(\mathfrak{h}_{1}+\frac{1}{2},\mathfrak{h}_{2}) \Big]z_{12}^{h_{p}+h_{1}-h_{2}}\bar{z}_{12}^{\bar{h}_{2}-\bar{h}_{p}-\bar{h}_{1}-1}L^{-}[\phi_{\mathfrak{h}_{p}+\frac{1}{2}}] \nn\\
  &i\epsilon \Big[\frac{D(\mathfrak{h}+\frac{1}{2},\mathfrak{h}_{2})}{1-2h_{1}}(h_{p}+h_{1}-h_{2})+D(\mathfrak{h}_{1}+\frac{1}{2},\mathfrak{h}_{2}) \Big]z_{12}^{h_{p}+h_{1}-h_{2}}\bar{z}_{12}^{\bar{h}_{2}-\bar{h}_{p}-\bar{h}_{1}}\partial_{\bar{z}_{2}}L^{-}[\phi_{\mathfrak{h}_{p}+\frac{1}{2}}] \nn\\
  &-i\epsilon \frac{D(\mathfrak{h}_{1},\mathfrak{h}_{2}+\frac{1}{2})}{1-2\bar{h}_{2}}z_{12}^{h_{p}+h_{1}-h_{2}-1}\bar{z}_{12}^{\bar{h}_{2}-\bar{h}_{p}-\bar{h}_{1}+1}z_{2}\partial^{2}_{\bar{z}_{2}}L^{-}[\phi_{\mathfrak{h}_{p}+\frac{1}{2}}]\nn \\
   &=-i\epsilon C(\mathfrak{h}_{1},\mathfrak{h}_{2})z_{12}^{h_{p}+h_{1}-h_{2}-1}\bar{z}_{12}^{\bar{h}_{2}-\bar{h}_{p}-\bar{h}_{1}}\bigg( \frac{z_{2}\partial_{\bar{z}_{2}}}{1-2\bar{h}_{p}}\bigg)L^{-}[\phi_{\mathfrak{h}_{p}+\frac{1}{2}}]\nn \\
   &-i\epsilon D(\mathfrak{h}_{1},\mathfrak{h}_{2})z_{12}^{h_{p}+h_{1}-h_{2}-1}\bar{z}_{12}^{\bar{h}_{2}-\bar{h}_{p}-\bar{h}_{1}+1}\bigg( \frac{z_{2}\partial_{\bar{z}_{2}}}{1-2\bar{h}_{p}}\bigg)\partial_{\bar{z}_{2}}L^{-}[\phi_{\mathfrak{h}_{p}+\frac{1}{2}}].\label{M10generalOPE}
\end{align}
Comparing the leading order terms on the left-hand side and right-hand side we get, apart from the ones already derived, equation \eqref{generalRecursion5}. Looking at the sub-leading terms (more specifically the fourth term in rectangular braces), we can find, by similar arguments as were given for the $\delta_{M_{01}}$ case, equation \eqref{generalRecursion6}.  

\section{Intermediate steps for checking translation symmetry of shadow OPE} \label{Intermediate steps for checking translation symmetry of shadow OPE}
Starting with \eqref{CollinearShadowOPECheck} and using the OPE of shadow gravitons, the left-hand side simplifies to 
\begin{align}
    &2i\epsilon \color{orange} \bigg[ \frac{\Gamma(\Delta_{p}-3)(\Delta_{p}+1)}{\Gamma(\Delta_{2}+1)\Gamma(\Delta_{3}-3)}-\frac{\Gamma(\Delta_{p}-2)(\Delta_{p}+1)}{\Gamma(\Delta_{2}+1)\Gamma(\Delta_{3}-2)}+\frac{\Gamma(\Delta_{p}-3)(\Delta_{p}+1)}{\Gamma(\Delta_{2})\Gamma(\Delta_{3}-2)}\bigg]\color{black}\frac{P(\Delta_{2},\bar{z}_{2},\bar{z}_{3})}{\mathrm{sin}(\pi\Delta_{2})\bar{z}_{23}^{3}}\partial_{z_{3}}S[G^{-}_{\Delta_{p}+1}]\nn \\
    &+i\epsilon \color{violet} \bigg[-\frac{\Gamma(2-\Delta_{2})\Gamma(\Delta_{p}-3)\Gamma(\Delta_{2})}{\pi \Gamma(\Delta_{2}+1)\Gamma(\Delta_{3}-3)}-\frac{\Gamma(3-\Delta_{2})\Gamma(\Delta_{p}-2)\Gamma(\Delta_{2}-1)}{\pi \Gamma(\Delta_{2}+1)\Gamma(\Delta_{3}-2)}\nn\\
    &\color{violet}+\frac{\Gamma(3-\Delta_{2})\Gamma(\Delta_{p}-3)\Gamma(\Delta_{2}-1)}{\pi \Gamma(\Delta_{2})\Gamma(\Delta_{3}-2)} -\frac{\Gamma(\Delta_{p}-2)(\Delta_{p}+1)}{\Gamma(\Delta_{2}+1)\Gamma(\Delta_{3}-2)\mathrm{sin}(\pi\Delta_{2})}\nn \\
    &\color{violet}+\frac{\Gamma(\Delta_{p}-3)(\Delta_{p}+1)}{\Gamma(\Delta_{2})\Gamma(\Delta_{3}-2)\mathrm{sin}(\pi\Delta_{2})}\bigg]\color{black}\frac{P(\Delta_{2},\bar{z}_{2},\bar{z}_{3})}{\bar{z}_{23}^{2}}\partial_{z_{3}}\partial_{\bar{z}_{3}} S[G^{-}_{\Delta_{p}+1}]\nn \\
    &-2i\epsilon \frac{\Gamma(\Delta_{p}-3)(\Delta_{p}+1)}{\Gamma(\Delta_{2})\Gamma(\Delta_{3}-2)\mathrm{sin}(\pi\Delta_{2})}\frac{P(\Delta_{2},\bar{z}_{2},\bar{z}_{3})z_{23}}{\bar{z}_{23}^{2}}\partial_{z_{3}} S[G^{-}_{\Delta_{p}+1}]\nn \\
    &-i\epsilon \bigg[\frac{\Gamma(\Delta_{p}-2)(\Delta_{p}+1)}{\Gamma(\Delta_{2})\Gamma(\Delta_{3}-2)\mathrm{sin}(\pi \Delta_{2})}+\frac{\Gamma(3-\Delta_{2})\Gamma(\Delta_{p}-3)\Gamma(\Delta_{2}-1)}{\pi\Gamma(\Delta_{2})\Gamma(\Delta_{3}-2)}  \bigg]\frac{P(\Delta_{2},\bar{z}_{2},\bar{z}_{3})z_{23}}{\bar{z}_{23}^{2}}\partial_{z_{3}}^{2}\partial_{\bar{z}_{3}}S[G^{-}_{\Delta_{p}+1}]\nn \\
    & +i\epsilon \color{blue} \bigg[ -\frac{\Gamma(3-\Delta_{2})\Gamma(\Delta_{p}-2)\Gamma(\Delta_{2}-1)}{\pi \Gamma(\Delta_{2}+1)\Gamma(\Delta_{3}-2)}+\frac{\Gamma(3-\Delta_{2})\Gamma(\Delta_{p}-3)\Gamma(\Delta_{2}-1)}{\pi \Gamma(\Delta_{2})\Gamma(\Delta_{3}-2)}\bigg]\color{black}\frac{P(\Delta_{2},\bar{z}_{2},\bar{z}_{3})}{\bar{z}_{23}}\partial_{z_{3}}\partial_{\bar{z}_{3}}^{2}S[G^{-}_{\Delta_{p}+1}]\nn \\
    &-i\epsilon \frac{\Gamma(3-\Delta_{2})\Gamma(\Delta_{p}-3)\Gamma(\Delta_{2}-1)}{\pi\Gamma(\Delta_{2})\Gamma(\Delta_{3}-2)}\frac{P(\Delta_{2},\bar{z}_{2},\bar{z}_{3})z_{23}}{\bar{z}_{23}}\partial_{z_{3}}^{2}\partial_{\bar{z}_{3}}^{2}S[G^{-}_{\Delta_{p}+1}].
\end{align}
where we have chosen $-\epsilon_{1}=\epsilon_{2}=\epsilon_{p}=\epsilon$ and have written the terms relevant for symmetry analysis in coloured text. The term in orange evaluates to,
\begin{align}
     \color{orange}\frac{\Gamma(\Delta_{p}-3)(\Delta_{p}+1)}{\Gamma(\Delta_{2}+1)\Gamma(\Delta_{3}-3)}-\frac{\Gamma(\Delta_{p}-2)(\Delta_{p}+1)}{\Gamma(\Delta_{2}+1)\Gamma(\Delta_{3}-2)}+\frac{\Gamma(\Delta_{p}-3)(\Delta_{p}+1)}{\Gamma(\Delta_{2})\Gamma(\Delta_{3}-2)}=\color{black}0.
\end{align}
The term in violet simplifies to, 
\begin{align}
    &\color{violet}-\frac{\Gamma(2-\Delta_{2})\Gamma(\Delta_{p}-3)\Gamma(\Delta_{2})}{\pi \Gamma(\Delta_{2}+1)\Gamma(\Delta_{3}-3)}-\frac{\Gamma(3-\Delta_{2})\Gamma(\Delta_{p}-2)\Gamma(\Delta_{2}-1)}{\pi \Gamma(\Delta_{2}+1)\Gamma(\Delta_{3}-2)}+\frac{\Gamma(3-\Delta_{2})\Gamma(\Delta_{p}-3)\Gamma(\Delta_{2}-1)}{\pi \Gamma(\Delta_{2})\Gamma(\Delta_{3}-2)}\nn\\
    &\color{violet} -\frac{\Gamma(\Delta_{p}-2)(\Delta_{p}+1)}{\Gamma(\Delta_{2}+1)\Gamma(\Delta_{3}-2)\mathrm{sin}(\pi\Delta_{2})}+\frac{\Gamma(\Delta_{p}-3)(\Delta_{p}+1)}{\Gamma(\Delta_{2})\Gamma(\Delta_{3}-2)\mathrm{sin}(\pi\Delta_{2})}=\color{black}-\frac{\Gamma(\Delta_{p}-3)\Delta_{p}}{\Gamma(\Delta_{2}+1)\Gamma(\Delta_{3}-3)\mathrm{sin}(\pi\Delta_{2})}.
\end{align}
Lastly, the terms in blue sum up to,
\begin{align}
    \color{blue}-\frac{\Gamma(3-\Delta_{2})\Gamma(\Delta_{p}-2)\Gamma(\Delta_{2}-1)}{\pi \Gamma(\Delta_{2}+1)\Gamma(\Delta_{3}-2)}+\frac{\Gamma(3-\Delta_{2})\Gamma(\Delta_{p}-3)\Gamma(\Delta_{2}-1)}{\pi \Gamma(\Delta_{2})\Gamma(\Delta_{3}-2)}=\color{black}-\frac{\Gamma(3-\Delta_{2})\Gamma(\Delta_{p}-3)\Gamma(\Delta_{2}-1)}{\pi \Gamma(\Delta_{2}+1)\Gamma(\Delta_{3}-3)}.
\end{align}
Now, using this one can write down \eqref{LHSShadowCollienarOPEM00Check}.
\bibliographystyle{JHEP}
\bibliography{main.bib}
\end{document}